\RequirePackage{fix-cm}
\documentclass[smallextended]{svjour3}       
\smartqed  
\usepackage{graphicx}
\usepackage{epstopdf}
\usepackage{natbib}
\usepackage{hyperref}
\usepackage{epsf}
\usepackage{epsfig}
\usepackage{multirow}
\usepackage[english]{babel}
\usepackage[latin1]{inputenc}
\usepackage{graphicx}
\usepackage{geometry}
\usepackage{subfigure}
\usepackage[T1]{fontenc}
\usepackage{lmodern}

\usepackage{amssymb}
\usepackage{amsmath}
\usepackage{color}
\usepackage{enumerate}

\newcommand{\btau}{\mbox{\boldmath{$\tau$}}}

\newcommand{\bmu}{\mbox{\boldmath{$\mu$}}}
\newcommand{\btheta}{\mbox{\boldmath{$\theta$}}}

\journalname{Computational Statistics}
\begin{document}

\title{Improved model-based clustering performance using Bayesian initialization averaging.
}

\titlerunning{Model-based clustering with BIA}        

\author{Adrian O'Hagan         \and
        Arthur White 
}


\institute{A. O'Hagan \at
              School of Mathematics and Statistics and The Insight Centre for Data Analytics, University College Dublin, Ireland \\
              Tel.: +353-1716-2428\\
              Fax: N/A\\
              \email{adrian.ohagan@ucd.ie}           
           \and
           A. White \at
              School of Computer Science and Statistics, Lloyd Institute, Trinity College Dublin, Ireland
}

\date{Received: June 26th 2016 / Accepted: TBD}

\maketitle

\begin{abstract}
The Expectation-Maximization (EM) algorithm is a commonly used method for finding the maximum likelihood estimates of the parameters in a mixture model via coordinate ascent. A serious pitfall with the algorithm is that in the case of multimodal likelihood functions, it can get trapped at a local maximum. This problem often occurs when sub-optimal starting values are used to initialize the algorithm. Bayesian initialization averaging (BIA) is proposed as an ensemble method to generate high quality starting values for the EM algorithm. Competing sets of trial starting values are combined as a weighted average, which is then used as the starting position for a full EM run. The method can also be extended to variational Bayes (VB) methods, a class of algorithm similar to EM that is based on an approximation of the model posterior. The BIA method is demonstrated on real continuous, categorical and network data sets, and the convergent log-likelihoods and associated clustering solutions presented. These compare favorably with the output produced using competing initialization methods such as random starts, hierarchical clustering and deterministic annealing, with the highest available maximum likelihood estimates obtained in a higher percentage of cases, at reasonable computational cost. For the Stochastic Block Model for network data promising results are demonstrated even when the likelihood is unavailable. The implications of the different clustering solutions obtained by local maxima are also discussed.
\keywords{Bayesian model averaging \and Expectation-maximization algorithm \and Finite mixture models
\and Hierarchical clustering \and Model-based clustering \and Multimodal likelihood.}
\end{abstract}

\section{Introduction}
\label{section1_introduction}

The Expectation-Maximization (EM) algorithm is a commonly used coordinate ascent method for finding the maximum likelihood estimates for the parameters in a mixture model \citep{dempster77}.
To initialize the algorithm, starting values are chosen for the allocations of observations to the available number of clusters and these lead to the initial calculation of the model parameters. Often this starting allocation is done on a random basis. However, once these parameter values have been specified, the algorithm proceeds in an entirely deterministic fashion.

An advantage of the EM algorithm is that it does not produce decreases in the likelihood function. A drawback of the algorithm is that it can converge to or get trapped for many iterations at a local maximum, leading to failure to reach the global maximum and resulting in an inferior clustering solution. \citet{zhou2010} describe this multimodality of the likelihood function as "one of the curses of statistics". Various convergence criteria can be used to stop the algorithm. However these criteria tend to characterize a lack of progress of the algorithm, rather than the fact that the global maximum has been reached.

Many adaptations to the EM algorithm have been proposed that address this issue by altering its form. These include, among others, an EM algorithm featuring maximization using the Newton-Raphson method \citep{redner_walker84}; the Classification EM (CEM) algorithm \citep{biernacki2003, Celeux_Govaert_1992}; the Moment Matching EM algorithm \citep{karlis2003}; the Annealing EM algorithm \citep{zhou2010}; a
hybrid EM/Gauss-Newton algorithm \citep{aitkin96}; the Expectation Conditional Maximization (ECM) algorithm
\citep{Meng_Rubin_92}; the emEM algorithm \citep{biernacki2003}; the Multicycle EM algorithm \citep{meng1993} and the Sparse EM algorithm \citep{neal1998}.

Alternatively this problem can be addressed by identifying high quality, as opposed to random, starting values for the algorithm in an attempt to optimize the eventual clustering solution. Recent developments of this approach include \citet{OHagan2012}, who develop a pyramid burn-in scheme, and \citet{Baudry_2015}, who propose a recursive initialisation strategy based on splitting a selected component from a previous clustering solution, in particular with a view to performing robust model selection.

Bayesian initialisation averaging (BIA) is proposed as a new and complementary method in this spirit. To generate starting values using BIA, a trial number of E-steps and M-steps are run on each of a sequence of random starting positions to give an updated sequence of potential starting positions. Using the likelihood associated with each updated starting position, a corresponding weight is calculated and a new single overall weighted average starting position is formed. The EM algorithm is then run to convergence from this juncture. 

For some types of model, the log-likelihood is intractable, and inference using EM is not possible. Our approach is demonstrated, in a Bayesian setting, for one such model using variational Bayes (VB) methods, an EM-type algorithm based on an approximation of the model posterior. Related algorithms for this type of scenario include iterated conditional models \citep{Besag1986}, the Broyden-Fletcher-Goldfarb-Shanno (BFGS) algorithm \citep{Byrd1995,Zhu1997} and sequential Monte Carlo for VB \citep{mcgrory2014}. The latter shares some features in common with annealing and pyramid burn-in.

\begin{figure}[tbp]
\begin{center}
\includegraphics[scale=0.8]{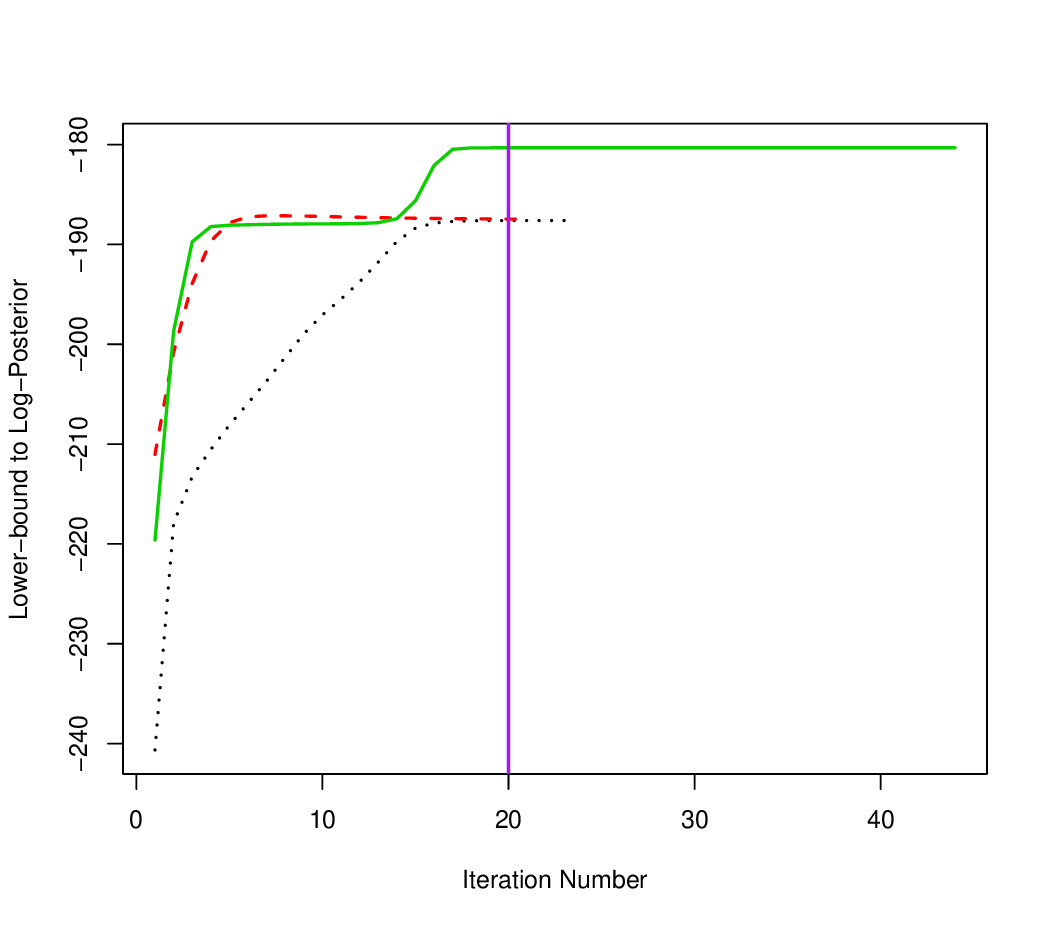}
\caption{Illustrative example of the Bayesian initialisation averaging approach applied to a clustering model for network data using an EM-like algorithm. The dashed lines show the value of the lower bound to the log-posterior at each iteration using suboptimal starting positions; by using BIA, a superior clustering solution (the solid line) is obtained.}
\label{fig:Karate_example}
\end{center}
\end{figure}

As an illustrative example of BIA, consider Figure $\ref{fig:Karate_example}$, which is based on a clustering model using a VB algorithm. (See Sections $\ref{section2 Karate}$ and $\ref{sec_mbc_examples}$ for specific details). The dashed lines show the value of the (lower bound to the) log-posterior at each iteration using two sets of randomly generated suboptimal starting positions. Using these starting positions, the algorithm converges to distinct solutions which are similarly valued in terms of log-posterior. Using BIA, it is possible to obtain a new, superior set of starting values, based on a weighted combination of the value of the two positions after running the algorithm for just $20$ iterations. Hence a superior clustering solution in terms of log-posterior (the solid line) may be obtained.

BIA starting values are used as an initialization method to cluster a variety of continuous, categorical and network data examples using Mixtures of Gaussians, Latent Class Analysis (LCA) and the Stochastic Block Model (SBM), implemented via the EM (Mixture of Gaussians and LCA) and VB (SBM). Even for the more complicated SBM approach where the likelihood can only be approximated, the proposed BIA method fares well. The approach is demonstrated on real and simulated data sets in comparison to competing methodologies such as random starts, hierarchical clustering, pyramid burn-in schemes and deterministic annealing.

The rest of the paper is organized as follows: Section $\ref{section2_data}$ presents the data sets used as motivating examples to illustrate the benefits of the proposed method.
Section $\ref{section3_EM_algorithm}$ gives a overview of how the EM algorithm is used for fitting mixture distributions. The calculations for the E-step and M-step are provided for different types of application.
A description of how to apply BIA to generate EM algorithm starting values is detailed. A label switching problem that is encountered, as well as a means of overcoming it, are documented.
In Section $\ref{section4_results}$ the outcomes of the methods tested on the data sets referenced in Section \ref{section2_data} are described. Convergent log-likelihood plots and clustering solutions are presented for BIA and its competing methodologies in each case.
Section $\ref{section5_conclusions}$ details the conclusions reached and summarizes the main findings.

\section{Illustrative Data Sets}
\label{section2_data}

Three types of data set are used as motivating examples in testing the BIA method. The first type is multivariate continuous data (the \emph{AIS} data set and a simulated data set taken from \cite{Baudry2015}); the second type is multivariate categorical data (two further simulated data sets and the \emph{Carcinoma} and \emph{Alzheimer's} data sets); the third type is network data (the \emph{Karate} data set). These are described below.

\subsection{Multivariate continuous data: Australian Institute of Sport (AIS) data and Simulated~Data~1}
\label{section2_AIS}

\begin{figure}[tbp]
\begin{center}
\includegraphics[scale=0.8]{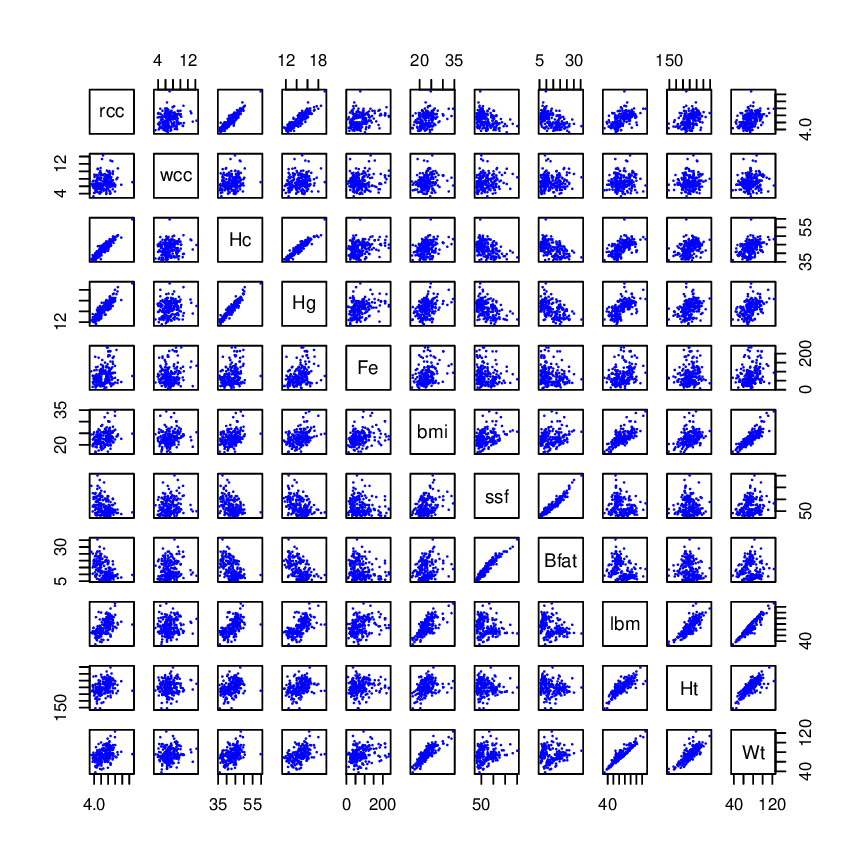}
\caption{Pairs plot of the \emph{Australian Institute of Sport} (\emph{AIS}) data.}
\label{fig:AIS_data}
\end{center}
\end{figure}

The \emph{Australian Institute of Sport} data set contains biometric observations on $n = 202$ Australian athletes ($102$ males and $100$ females) at the Australian Institute of Sport \citep{Cook_Weisberg_1994}. The full data set contains $13$ variables ($11$ continuous and $2$ discrete) and all continuous variables are included in the analysis: red cell count (rcc), white cell count (wcc), hematocrit (Hc), hemoglobin (Hg), plasma ferritin (Fe), body mass index (bmi), sum of skin folds (ssf), body fat percentage (Bfat), lean body mass (lbm), height (Ht) and weight (Wt) \citep{Cook_Weisberg_1994} (see Figure $\ref{fig:AIS_data}$).

\begin{figure}[tbp]
\begin{center}
\includegraphics[scale=0.8]{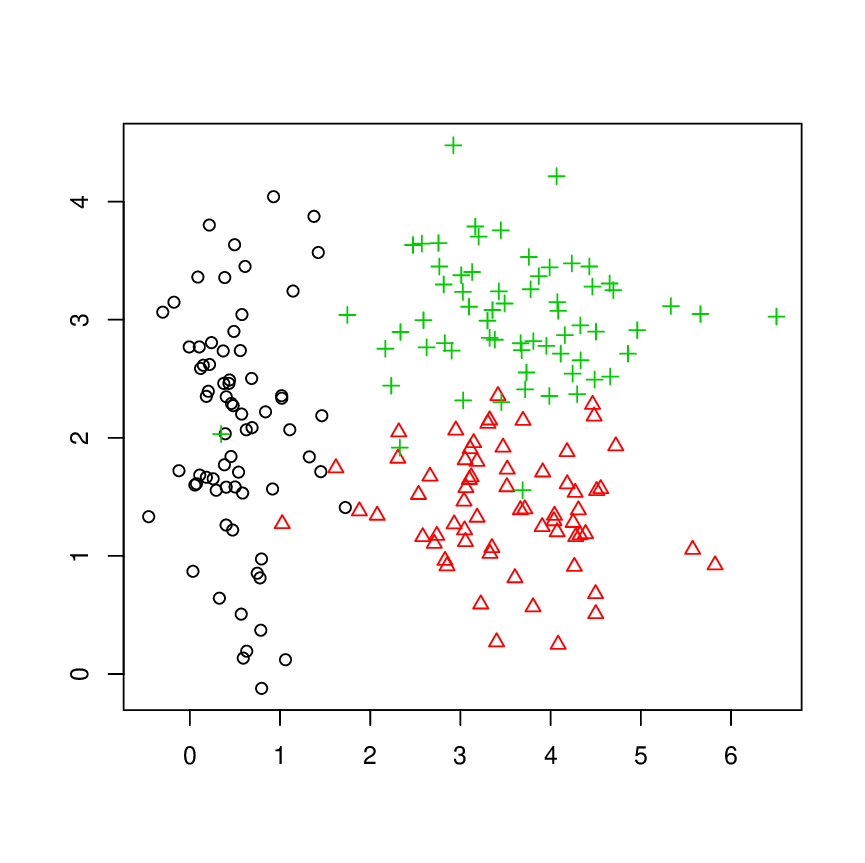}
\caption{Pairs plot of one instance of the simulated mixture of Gaussians data (Simulated~Data~$1$ $n=200$) taken from \cite{Baudry2015} experiment $3$.}
\label{fig:BadurySimData}
\end{center}
\end{figure}

A further simulated data set, Simulated Data $1$, taken from \cite{Baudry2015} experiment $3$ is tested, namely $n = 200$ observations simulated from a mixture of three multivariate Gaussian components with diagonal covariance structure. The second and third horizontal components (plotted as red triangles and green crosses respectively) overlap, giving the appearance of forming a single cluster, as seen in Figure $\ref{fig:BadurySimData}$. The first component, plotted as black circles, is vertical in nature. For this data set, both a $2$ group and a $3$ group solution are relevant. $100$ such data sets are randomly simulated for the purposes of the study. The process is repeated to create variants of Simulated Data $1$ with larger sample sizes, specifically for $n = 500, n = 1000$ and $n = 5000$.

\subsection{Multivariate categorical data: Simulated~Data~2, Simulated Data 3, Carcinoma, and Alzheimer's data}
\label{section2 Carcinoma}
\begin{figure}[tbp]
\begin{center}
\includegraphics[scale=0.95]{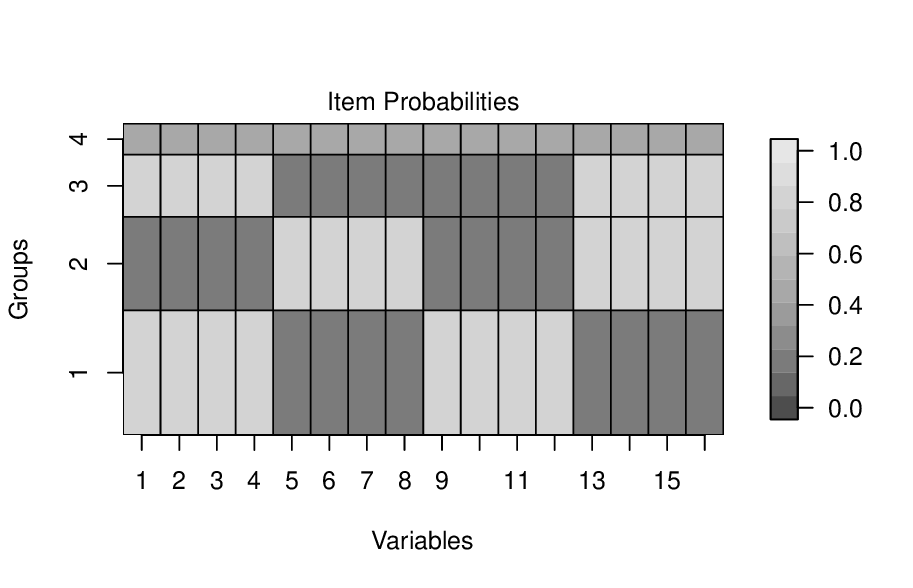}
\caption{Visualisation of model parameters used for Simulated Data $2$.}
\label{fig:simstudy2}
\end{center}
\end{figure}

To assess the performance of the proposed BIA approach to  different procedures (described in Section~\ref{section3_sub_hclust}), a simulation study is conducted. Data are generated from a $4$ group LCA model (Simulated Data 2). An overview of the LCA model is provided in Section~\ref{section3_sub_fitting_models}. The data have $16$ variables, and were generated for a range of sample sizes, $n = \lbrace 100, 250, 500, 1000, 5000\rbrace.$ The two largest groups in this study are reasonably distinct, while the third group resembles the first group for the first $8$ data variables, and the second group for the latter $8$ variables. The fourth group is the smallest, and has the parameters with highest variance. See Table~\ref{tab:simstudy2} for a full specification of the model parameters, and Figure~\ref{fig:simstudy2} for a visualisation. These represent the item probability parameters as a heat map, with values close to $1$ represented by a lighter grey colour, and values close to $0$ with darker grey. The class size is represented by the relative length of the cells in each row.

A further simulation study for a $4$ group LCA model (Simulated Data 3) was conducted for the largest sample sizes $n = \lbrace 1000, 5000\rbrace.$ In this case the group weights were highly unbalanced, so that $\boldsymbol \tau = (0.52, 0.42, 0.05, 0.01)$. The item probability parameters were kept the same as in Table~\ref{tab:simstudy2}.

\begin{table}[bt]
\centering
\caption{Model parameters for Simulated Data $2$ simulation study.}
\begin{tabular}{ll}
Group weight & Item probability \\
  \hline
$\tau_1 = 0.4$ & $
\theta_{1m} = \left\{ \begin{array}{ll}
0.8, & m = 1, \ldots, 4, 9, \ldots, 12; \\
0.2 & \mbox{otherwise}.
\end{array} \right.$ \\
$\tau_2 = 0.3$ & $
\theta_{2m} = \left\{ \begin{array}{ll}
0.2, & m = 1, \ldots, 4, 9, \ldots, 12; \\
0.8 & \mbox{otherwise}.
\end{array} \right.$ \\
$\tau_3 = 0.2$ & $
\theta_{3m} = \left\{ \begin{array}{ll}
0.8, & m = 1, \ldots, 4, 13, \ldots, 16; \\
0.2 & \mbox{otherwise}.
\end{array} \right.$ \\
$\tau_4 = 0.1$ & $
\theta_{4m} = 0.5, m = 1, \ldots, 16.$
\end{tabular}
\label{tab:simstudy2}
\end{table}

Both the \emph{Carcinoma} and \emph{Alzheimer's} data sets consist of multivariate binary data, to which LCA is applied. The \emph{Carcinoma} data set contains dichotomous ratings by $m = 7$ pathologists of $n = 118$ slides for the presence or absence of carcinoma in the uterine cervix.
The data set is available in the \textbf{R} package \textbf{poLCA} \citep{poLCA} and is described at length by \citet[Section 13.2]{agresti2002}. As per the simple percentage diagnosed summary for each doctor in Table $\ref{tab:carc_data}$, the data clearly demonstrate that some pathologists are more prone to diagnosing the presence of carcinoma than others. Using LCA the aim is to identify sets of observations for which they agree and disagree. The approach of  \citet{zhou2010} is followed and the case of $G = 4$ classes considered.

\begin{table}[ht]
\caption{Percentage of slides identified as having cancer present by the seven different doctors in the \emph{Carcinoma} study.}
\centering
\begin{tabular}{lrrrrrrr}
\hline
Pathologist & A & B & C & D & E & F & G \\
\hline
Detected (\%)  & \,\,0.56 & \,\,0.67 & \,\,0.38 & \,\,0.27 & \,\,0.60 & \,\,0.21 & \,\,0.56 \\
\hline
\end{tabular}
\label{tab:carc_data}
\end{table}

The \emph{Alzheimer's} data set documents the presence or absence of $m = 6$ symptoms in $n = 240$ patients diagnosed with early onset Alzheimer's disease, conducted in Saint James' Hospital, Dublin, Ireland. This data is available in the \textbf{R} package  \textbf{BayesLCA} \citep{BayesLCA},  and was previously analyzed by \cite{Walsh04} and \cite{Walsh06}, where the goal was to identify whether groups of patients diagnosed with early onset Alzheimer's disease presented distinctive patterns of symptoms. The case where $G = 3$ latent classes are assumed to be present in the data is investigated.

\subsection{Network data: Karate data}
\label{section2 Karate}
\begin{figure}[tbp]
\begin{center}
\includegraphics[scale=0.5]{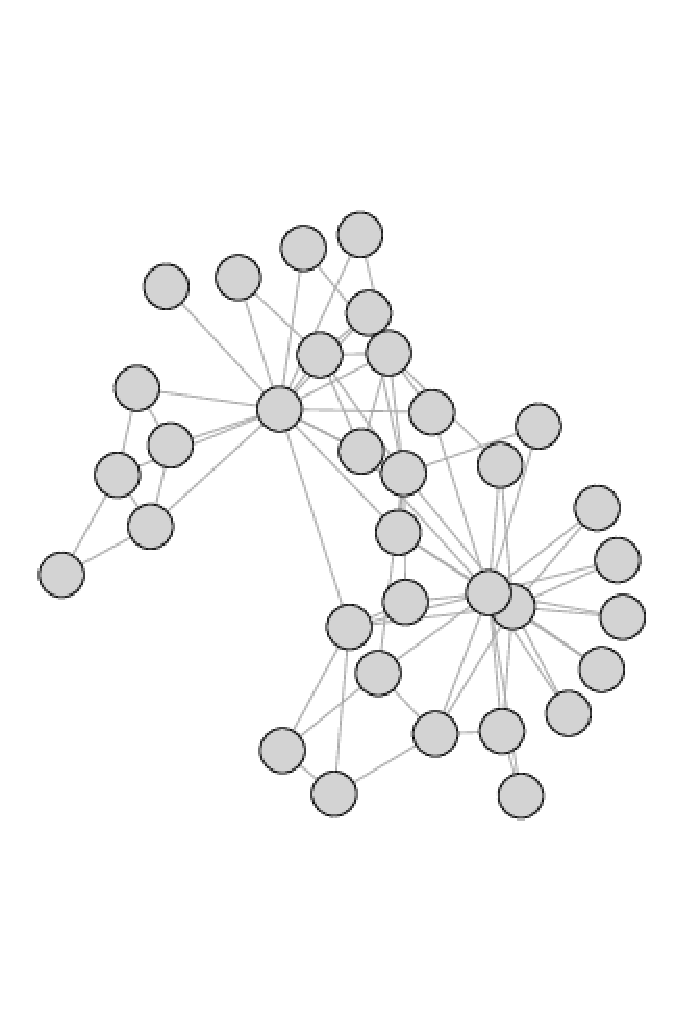}
\caption{Plot of the \emph{Karate} network data set.}
\label{fig:Karate_data}
\end{center}
\end{figure}

Zachary's Karate club data set records the social network of friendships between $n = 34$ members of a karate club at a United States university in the $1970$s \citep{zachary77}. While friendships within the network initially arose organically, a dispute between a part time instructor and the club president led to two political factions developing, with the members taking the side of either the president or the instructor, ultimately leading to the formation of a new organisation. The goal of the study is to correctly classify each member as ultimately taking the side of either the president or the instructor after the split. A $G = 4$ group SBM is applied to this data. The network is visualised in Figure $\ref{fig:Karate_data}$ using the Fruchterman-Reingold \citep{Fruchterman1991} layout algorithm available in the \textbf{R} package \textbf{igraph} \citep{igraph}.

\section{Methods}
\label{section3_EM_algorithm}

\subsection{The EM Algorithm}
\label{section3_sub_EM}

The EM algorithm is a powerful computational technique for finding the maximum likelihood estimates (MLEs) of the parameters in a mixture model when there are no closed-form maximum likelihood estimates, or when the data are incomplete \citep{McLachlan_Krishnan1997}. The EM algorithm was introduced by \cite{dempster77} and has many different applications, including cluster analysis, censored data modelling, mixed models and factor analysis. In model-based clustering, the EM algorithm maximizes the expected complete data log-likelihood to produce updated parameter values. Ultimately this process maximizes the observed data log-likelihood. The algorithm involves two steps, the E or expectation step and the M or maximization step. Initial starting values are required for the model parameters and the E and M steps are then iterated until convergence. In terms of initializing the algorithm the aim is to generate good-quality starting values that will lead to the algorithm converging to the highest possible log-likelihood.

\subsection{Fitting mixture models using the EM Algorithm}
\label{section3_sub_fitting_models}
Let $\mathbf{X}$ be the matrix of all $n$ observations and the $m-$dimensional vector $\mathbf{x}_i = ({x}_{i1}, \ldots, {x}_{im})$ denote the value of the $i^{th}$ observation, for $i = 1,2,...,n$. Each observation belongs to one of $G$ groups. $\mathbf{Z} = (z_{ig})$ is an $(n \times G)$
classification matrix for the matrix $\mathbf{X}$, where $\mathbf{z}_{ig} = 1$ when observation $i$ belongs to group $g$ and $\mathbf{z}_{ig} = 0$ otherwise. Group labels are not known in advance. $\mathbf{Z}$ is an $(n \times G)$ classification matrix of $\mathbf{z}_{ig}$ values for the matrix $\mathbf{X}$. $\tau_{g}$ is the probability that $\mathbf{x}_i$ belongs to group $g$. $\btheta_{g}$ denotes the distributional parameters for group $g$. The composition of the parameter set $\btheta$ varies depending on the application at hand; for example, in a continuous data setting where a mixture of Gaussians is fitted, $\btheta$ could represent the group means and covariance matrices. Further examples are given in Section $\ref{sec_mbc_examples}$.

If each observation $\mathbf{x}_i$ belongs to one of G groups with probability $\tau_{g}$ then the density of $\mathbf{x}_i$, conditioning on the distributional parameters $\btheta$ and group membership probabilities $\btau$, is given by:
\begin{equation}
\displaystyle\sum_{g=1}^{G} \left[\tau_g\,P\left(\mathbf{x_i}|\btheta_g\right)\right]
\label{eq:density_of_xi}
\end{equation}
where $P\left(\mathbf{x_i}|\btheta_g\right)$ is the density of $x_i$ if observation $i$ belongs to group $G$.

The observed likelihood $L$ is the product of the densities of the $\mathbf{x}_i$, conditioning on $\btheta$ and $\btau$:

\begin{equation}
L(\btau,\btheta) =  P(\mathbf{X}|\btau,\btheta) = \displaystyle \prod_{i=1}^{n}\displaystyle\sum_{g=1}^{G} \left[\tau_g\,P\left(\mathbf{x_i}|\btheta_g\right)\right].
\label{eq:observed_likelihood}
\end{equation}
This function is difficult to work with as its natural logarithm cannot be differentiated in a straightforward manner. However, introducing the $\mathbf{Z}$ classification matrix, the complete data likelihood $L_c$ can be expressed as:
\begin{equation}
L_c(\btau, \btheta, \mathbf{Z}) = P(\mathbf{X}, \mathbf{Z}|\btau,\btheta) = \displaystyle \prod_{i=1}^{n}\displaystyle\prod_{g=1}^{G} \left[\tau_g\,P\left(\mathbf{x_i}|\btheta_g\right)\right]^{z_{ig}}.
\label{complete_data_likelihood}
\end{equation}
The complete data log-likelihood $l_c$ can then be written as:
\begin{equation}
l_c(\btau, \btheta, \mathbf{Z}) = \displaystyle \sum_{i=1}^n\sum_{g=1}^G z_{ig}\left[\log \tau_g + \log P\left(\mathbf{x_i}|\btheta_g\right)\right].
\label{eq:complete_data_log_likelihood}
\end{equation}
The EM algorithm maximizes the observed log-likelihood, ($\ref{eq:observed_likelihood}$), according to the following process:

\begin{enumerate}[(i)]
\item
\textbf{Starting Values}:\,\,\,Set $\mathbf{Z}^{(0)}$ at iteration $t=0$. Calculate the values of $\btau^{(0)}$ and $\btheta^{(0)}$ based on $\mathbf{Z}^{(0)}$.
\vspace{5mm}
\item
\textbf{E-Step}: Evaluate $Q^{(t+1)} = E\left[l_c(\btheta,\btau,\mathbf{Z})|\theta^{(t)}, \btau^{(t)}\right]$, which means estimating $E(z_{ig})$:

\begin{equation}
\hat{e}_{ig}^{(t+1)}=E\left[z_{ig}^{(t+1)}\right]=\displaystyle\frac{\tau_g^{(t)}P(\mathbf{x}_i|\btheta_g^{(t)})}
{\displaystyle\sum_{g'=1}^G\tau_{g'}^{(t)}P(\mathbf{x}_i|\btheta_{g'}^{(t)})}.
\label{eq:z}
\end{equation}

\item
\textbf{M-Step}:\,\,\,Maximize $Q^{(t+1)}$ (Equation $\ref{eq:Q}$) with respect to the group membership probabilities, $\btau$, and the distributional parameters of the observed data, $\btheta$. This produces new values $\btau^{(t+1)}$ (see Equation $\ref{eq:tau}$) and $\btheta^{(t+1)}$, where the composition of $\btheta^{(t+1)}$ varies according to the type of data:

\begin{eqnarray}
Q^{(t+1)}&=&\displaystyle\sum_{i=1}^n\sum_{g=1}^G
\hat{e}_{ig}^{(t+1)} \left[\log\tau_g+\log
P(\mathbf{x}_i|\btheta_g)\right]
\label{eq:Q}\\
\hat{\tau}_g^{(t+1)}&=&\displaystyle\frac{1}{n}\displaystyle\sum_{i=1}^n\,\hat{e}_{ig}^{(t+1)}.
\label{eq:tau}
\end{eqnarray}

\item
\textbf{Convergence Check}: The algorithm is stopped when the observed log-likelihood has converged. Convergence is deemed to have been reached when the relative change in the log-likelihood is ``sufficiently small'':

\begin{equation}
\displaystyle\frac{l(\btheta^{(t+1)},\btau^{(t+1)}) -
l(\btheta^{(t)},\btau^{(t)})}
{l(\btheta^{(t+1)},\btau^{(t+1)})} < \epsilon,
\label{eq:relative_change_loglik}
\end{equation}
where $l(\theta,\tau) = \log L(\theta,\tau)$ and
$\epsilon$ is a suitably small value specified by the
user. For the continuous data and Stochastic Block Model applications (\emph{AIS} and \emph{Karate} data respectively), $\epsilon$ was chosen as $1 \times 10^{-5}$ to keep the method consistent with alternative approaches such as hierarchical clustering initialization in \textbf{mclust} \citep{fraley99}, which use the same convergence criterion. For the Latent Class Analysis examples (\emph{Carcinoma} and \emph{Alzheimer's} data), $\epsilon$ was chosen to be $1 \times 10^{-9}$, in keeping with previous experiments performed by \citet{zhou2010}. The EM algorithm parameter estimates $\hat{e}_{ig}$ at convergence are used as a means of clustering the data set. Observation $i$ is clustered to belong to the group $g$ that produces the largest value of $\hat{e}_{ig}$.
\end{enumerate}

A common problem with the EM algorithm is that it can get trapped at a local maximum, leading to an inferior clustering solution. A computationally intensive remedy is to re-run the algorithm from many random $\mathbf{Z}$ starting positions to convergence and then select the solution with the highest log-likelihood. Other approaches include specifying starting values by first using some type of deterministic heuristic algorithm, or employing an annealing approach. Alternatively, it is possible to optimize starting values by combining multiple random starts together. Bayesian initialization averaging (BIA)  provides a means of achieving this goal by using a weighted combination of $\mathbf{Z}$ starting positions.

While the objective is generally to find the global log-likelihood maximum, it must be noted that this mode may not be optimal in terms of how meaningful the clustering is. In models where there are no restrictions on the covariance structure the maximum log-likelihood is actually infinite. In theory, the method could result in an infinite likelihood if one of the candidate Z matrices attains a clustering solution corresponding to such a likelihood over the course of the preliminary iterations carried out. However, the restriction to a relatively small number of preliminary runs for each candidate $\mathbf{Z}$, and the fact that many $\mathbf{Z}$ matrices are averaged before the algorithm is run to convergence, helps to render this less likely than when using random starting initialisations. This constitutes a benefit of the proposed methodology since infinite likelihoods typically correspond to problematic clustering solutions and unstable parameter estimates.

Additionally, some spurious solutions will yield convergent log-likelihoods that are ``too high'', typically in cases where additional components are fitted to capture small groups of outlying observations. \cite{McLachlanPeel2000} document this phenomenon for mixtures of normal distributions. For the \emph{AIS}, \emph{Carcinoma}, \emph{Alzheimer's} and \emph{Karate} data sets, spurious solutions are not an issue due to the number of clusters considered, which are in turn guided by previously published results under competing methodologies. Overall, any reference to the ``global mode'' should be interpreted as referring to the highest finite log-likelihood that emerges from a meaningful clustering solution in a stable subset of the parameter space (i.e. that the number of observations in any cluster and hence the cluster variance does not become extremely small in the context of the other cluster compositions). A log-likelihood mode is designated as a ``global mode'' if it is either equal to the maximum log-likelihood achieved under previous methodologies (if no higher log-likelihoods are identified); or the highest log-likelihood achieved (if log-likelihoods higher than the maximum previously identified are found); while the clustering solution remains meaningful and the parameter estimation stable.

On a related note, it is worth recognising that simply ranking models by convergent likelihood is not on its own proof of superior clustering fit and hence a superior initialization scheme. However, for the motivating data sets presented, the clustering solutions corresponding to higher convergent likelihoods at a very minimum give valuable insight as to alternative interpretations of how the data can be clustered; which can be more intuitive or insightful. This point is analyzed by \cite{McLachlanPeel2000}, who argue that for continuous data the inter-component sum of squared distances to the cluster mean tends to improve (decrease) as convergent log-likelihood increases.

\subsubsection{Examples of mixture models}
\label{sec_mbc_examples}

The following mixture models were used to cluster the data sets described in Section $\ref{section2_data}$.

\begin{description}
\item[Mixture of Gaussians] In the continuous data setting, where $\mathbf x_{i} \in {\mathbb R}^m,$ a mixture of Gaussians \citep{fraley99} can be fitted. Here $\btheta$ represents the group means $\bmu$ and the group variances/covariances $\boldsymbol \sigma$ or $\boldsymbol \Sigma$, depending on whether the data is univariate or multivariate respectively. Several possible covariance structures are available; see \citet{fraley99} for further details.
\item[Latent Class Analysis] LCA \citep{goodman1974} involves clustering of categorical (specifically nominal) data, which in the simplest case is binary in nature denoting, e.g., the presence or absence of a symptom. In this setting, each $x_{im} \in \lbrace 0, 1 \rbrace,$ and $\btheta$ is a $G \times M$ matrix often referred to as the item probability parameter, that is, $\theta_{gm}$ denotes the probability, conditional on membership of group $g$, that $x_{im} = 1$, for any $i \in 1, \ldots, n$,
\item[Stochastic Blockmodel] Social network analysis \citep{Wasserman1994} is the study of how links, such as friendship, between a set of actors are formed. Here, ${ \mathbf X}$ is  an $n \times n$ matrix such that \begin{equation*}
x_{ij} = \left\{ \begin{array}{ll}
1 & \mbox{if a link exists between actors $a_i$ and $a_j$} \\
0 & \mbox{otherwise}.
\end{array} \right. 
\end{equation*}
The SBM \citep{holland1983,Snijders1997,Daudin2008} introduces $G$ latent groups underlying the network. Here, $\btheta$ is a $G \times G$ connectivity matrix that, conditional on their group membership, represents the probability of a link being formed between two actors.
    Note that inference in this setting is not as straightforward as for the mixture of Gaussians and LCA cases, and the EM algorithm cannot be used directly for this model. This is because the observed data likelihood function is not available in closed form, and an approximation is required; see \cite{Daudin2008} for further details. \citet{volant2012} use a VB method to perform inference for the SBM. This approach introduces a tractable lower bound ${\mathcal L} < l$ to the observed data log-likelihood which can itself be maximised in an EM-like iterative manner. To make use of this method requires specifying prior distributions. In the example in Section $\ref{section4 sub Karate}$, uninformative (proper) priors were chosen.
\end{description}

The LCA and SBM approaches make a conditional independence assumption~"\citep{hand2001}, whereby, conditional on group membership, the data points are assumed to be independent. Mixture models also assume conditional independence - the covariance structure models dependence between variables as opposed to data points.

\subsection{Generating Starting Values}
\label{section3_sub_hclust}

\subsubsection{Random Allocation}

A simple method for initializing the EM algorithm is to randomly allocate observations to the available groups. In the examples in Section $\ref{section4_results}$, unless otherwise stated, the algorithm is initialized according to the following method: $\mathbf{z}^{(0)}_{i} \sim \mbox{Multinomial}(1, 1/G, \dots, 1/G), \mbox{ for each }i \in 1, \ldots, n.$

Alternatively, the algorithm can be initialized by specifying random values for the parameters $\btheta$ and $\btau$. For example, in a Bayesian setting, these could be generated from the model priors. However, following experimentation, superior performance was achieved by randomly generating values for $\mathbf{Z}^{(0)}$.

\subsubsection{Hierarchical Clustering Starting Values}
\label{section3_sub_hclust}

For multivariate continuous data sets, \textbf{mclust} utilizes hierarchical clustering to generate a starting $\mathbf{Z}$ value for the EM algorithm. Hierarchical clustering constructs a tree-like structure known as a dendrogram to show groups of observations. The final clustering is built up over a number of steps, where similar observations are joined together. Several measures can be used to calculate the distance or dissimilarity between observations, with Euclidean distance the most common. There are also many criteria of aggregation that may be used such as the single linkage criterion, the complete linkage criterion and Ward's method \citep{Ward}. The observations with the lowest dissimilarities are grouped together first and the process then iterates on the group level until all observations belong to a single group and the dendrogram mapping the process is constructed. Once the tree is formed, the number of groups identified as optimal depends on where the tree is cut. The dendrogram is conventionally cut where there is relatively wide range of distances over which the number of clusters in the solution does not change. An advantage of hierarchical clustering is that the number of clusters is not assumed in advance. However the starting values yielded are not necessarily optimal in terms of eventually achieving the global maximum log-likelihood. BIA represents an alternative to this initialization method for generating better-quality starting values to facilitate ultimately obtaining the highest possible convergent log-likelihood.

\subsubsection{Burn-in Methods and Parameter Targeting}
\label{section3_sub_burnin_targeting}

\cite{OHagan2012} detail two novel EM algorithm initialization schemes for model-based clustering of multivariate continuous data. The ``burn-in'' methods proposed start with a set of candidate $\mathbf{Z}$ matrices, run each $\mathbf{Z}$ for a number of EM iterations, rank the $\mathbf{Z}$s according to likelihood and retain only a certain fraction of the candidate $\mathbf{Z}$s for the next set of EM runs, until only one $\mathbf{Z}$ remains. The parameter targeting method evaluates which parameter(s) in the model are most effective in driving the likelihood uphill at consistent intervals and focuses M steps on those parameters, significantly reducing the overall computational burden. BIA represents a significantly faster and statistically more robust alternative to these schemes.

\subsubsection{Deterministic Annealing}
\label{section3_sub_tempering}

An alternative means of identifying a global likelihood maximum is provided by deterministic annealing \citep{ueda1998,zhou2010}. Under this approach, a positive tuning parameter $\nu$ is used to flatten the likelihood function, with the adapted likelihood rendered easier to explore and ultimately identify the global maximum using the EM algorithm. If implemented correctly, the algorithm should then obtain the global maximum regardless of the initial starting values.

In the case of finite mixture models, the approach is implemented by setting \[ \hat{e}_{ig}^{(t+1)}= \displaystyle\frac{ \left [ \tau_g^{(t)}P(\mathbf{x}_i|\btheta_g^{(t)}) \right ] ^{\nu^{(\kappa)}}}
{\displaystyle\sum_{g'=1}^G \left [ \tau_{g'}^{(t)}P(\mathbf{x}_i|\btheta_{g'}^{(t)}) \right ]^{\nu^{(\kappa)} }} , \]  where $\nu^{0}$ is initially set to some small value, e.g., $\nu^{0} = 0.05$, then increased every $s$ iterations towards $\nu^{\infty} = 1$. The positively valued $r$ determines the rate of convergence to $\nu^\infty$, by setting $\nu^{(\kappa+1)} = r \nu^{(\kappa)} + (1 - r)\nu^{\infty}.$  Note that $\nu^0, r$ and $\nu^{\infty}$ must all be specified by the user.

\subsection{Bayesian Initialization Averaging}
\label{section3_sub_BIA}

In standard statistical practice, a single model is selected from a collection of competing models as the "best" model, using a metric such as BIC, and inferences are made on the basis of this "best" model being the true model \citep{Raftery05}. Consequently the element of model uncertainty is often ignored. As a general framework, Bayesian model averaging (BMA) can provide a mechanism to account for this uncertainty \citep{Hoeting99}. By accounting for model uncertainty, BMA minimizes prediction risk and has also been shown to improve model prediction accuracy on average by \cite{Wintle2003}, \cite{Volinsky1997} and \cite{Murphy2001}, among others.
It does this by averaging over competing models to provide broader conclusions about parameter estimates than from using a single model. Usually BMA is used after models are fitted in order to weight over the convergent results of multiple models. Conversely, its use is now proposed as an initialization method to generate starting values for the EM algorithm in an attempt to maximize the convergent log-likelihood.

The rationale for using a BMA-like approach to initialization is as follows: by averaging candidate starting values according to their relative quality of fit at a pre-convergence stage in running the EM algorithm, the overall weighted starting value is then capable of matching or exceeding the performance of any single individual candidate starting value, as it combines information on the allocation of observations to clusters from multiple evolutions of the model-fitting process. From a computational perspective, the approach allows a large number of candidate solutions to be proposed and evaluated in an efficient manner.

The following steps are followed to apply Bayesian \emph{initialization} averaging (BIA) to generate the EM algorithm starting values for model-based clustering. In the case of the SBM, the lower bound ${\mathcal L}$ is used as a substitute for $l$.

\begin{enumerate}[(i)]

\item
Specify a model structure, as well as a set of candidate models to be used to generate starting values. While only a single model structure is considered for each of the motivating data sets presented, we present the more general case where several candidate models are used to generate starting values. For example, a set of candidate models could have differing covariance structures, or a different number of groups. Some of the practical elements of this latter approach are discussed in Section~\ref{section3_sub_label_switching}.

\item
Run a small number of preliminary E-steps and M-steps on each of a series of randomly generated $\mathbf{Z}$ starts to give an updated sequence of $\mathbf{Z}$ matrices, using the model structure(s) identified in (i).

\item
Calculate the value of an information criterion similar in nature to the Bayesian Information Criterion \citep{Schwarz1978} associated with each updated $\mathbf{Z}$ matrix. ${BIC^*_{Z_j}}$ is the value of this criterion associated with the $j^{th}$ $\mathbf{Z}$ matrix, calculated as:

\begin{equation}
BIC^*_{Z_j} = -2l_{Z_j} + p_j\log(n).
\label{BIC}
\end{equation}
where $l_{Z_j}$ is the observed log-likelihood for the model corresponding to the clustering solution represented by the $j^{th}$ $\mathbf{Z}$ matrix, $p_j$ is the number of parameters in the model corresponding to the clustering solution represented by the $j^{th}$ $\mathbf{Z}$ matrix and $n$ is the number of observations. The key difference between $BIC^*$ and the standard formulation is that the criterion is not evaluated at the maximum observed likelihood, since this quantity will not yet have been identified.

Despite not operating at the maximum observed likelihood, empirically it is found that the penalty term $p_j\log(n)$ remains an effective discriminator between models of varying complexity when assigning weights to the competing starting positions. The reason that the BIC is used, as opposed to alternatives such as AIC or the Hannan and Quin criterion, is that, by approximating the (log of) the Bayes factor, a straightforward approximation to averaging the posterior distribution under each $\mathbf{Z}$ matrix can be calculated~\citep{Hoeting99}. Additionally, the penalty for increased parameter count for BIC is higher than for AIC or Hannan Quin. This is often desirable when attempting to achieve model parsimony. The consistency property of the BIC (see, for example, \cite{Keribin2000}) means that as the sample size grows the inclusion threshold for increased model complexity (for example an increase in group count) becomes more stringent; with BIC increasingly isolating the $\mathbf{Z}$ matrices that exhibit the greatest potential for optimising the log-likelihood function.

\item
Re-scale the $BIC^*$ values by subtracting the maximum value from each of the individual elements. This renders the results computationally tractable for evaluation in the remaining steps.

\item
Calculate the weight for the $j^{th}$ $\mathbf{Z}$ matrix, $w_j$, for BIA using the formula:
\begin{equation}
w_{j} = \dfrac{\exp(-0.5 BIC^*_{Z_j})}{\sum_k{\exp(-0.5 BIC^*_{Z_k})}}.
\label{eq:BMA_weight}
\end{equation}

The reason each weight is calculated in this form is that it constitutes a straightforward approximation of posterior model probability based on the $\mathbf{Z_j}$ matrix in question, with the exponential weightings used to appropriately adjust the Bayes factor approximation of the BIC from the log scale. Hence this a straightforward implementation of Bayesian model averaging, assuming a non-informative prior distribution for the competing $\mathbf{Z}$ matrices. The construction of the weights based on BIC values confers all of the advantages detailed in step $(iii)$ of achieving a balance between model goodness of fit (higher likelihood) and model parsimony (fewer parameters).

\item
Form a single new $\mathbf{Z}$ matrix, $\mathbf{Z^{*}}$, as the weighted average of the $\mathbf{Z}_j$ matrices obtained at the end of step $(ii)$ using the values calculated by expression $\ref{eq:BMA_weight}$ as the weights:

\begin{equation}
\mathbf{Z^{*}} = \sum_{j}w_{j}\mathbf{Z}_{j}.
\label{eq:single_Z}
\end{equation}

\item
Run the EM algorithm to convergence using the single BIA weighted starting $\mathbf{Z^{*}}$ matrix formed in (vi).

\end{enumerate}

\subsection{The Label Switching Problem}
\label{section3_sub_label_switching}

A label switching problem is encountered when applying the BIA method, prior to implementing step (vi) above. The maximum likelihood label configurations from the runs of EM algorithm are invariant to permutations of the data labels. Hence, if the BIA weighting process is applied to a large number of candidate $\mathbf{Z}$ matrices that have undergone a preliminary set of EM steps, the result is merely a single $\mathbf{Z}$ matrix with roughly equal membership probabilities for all observations across all available groups.

The method employed to undo the label switching problem is a ``soft'' label switching method. This method involves using the actual $\mathbf{Z}$ values or "soft" group membership probabilities associated with each observation \citep{Slonim2005}. Hence observations on the boundaries between several clusters are not forced to fully belong to one of the clusters, but rather are assigned membership percentages indicating their partial membership \citep{Rokach}. The \textbf{matchClasses} function in the \textbf{R} package \textbf{e1071}\citep{e1071} was used to switch the labels.

The \textbf{matchClasses} function takes the actual values in the array of $\mathbf{Z}$ matrices and matches them with a fixed row in the $\mathbf{Z}$ matrix to fix the labels. Each row of the matrix is sequentially matched versus the fixed row until the overall labelling structure is repaired. Using the actual $\mathbf{Z}$ probabilities avoids the loss of information associated with label-switching methodologies that operate on the ``hard'' group labels, such as that associated with the work of \cite{Carpaneto1980} and \cite{Nobile07}. However, for completeness, the latter ``hard'' label-switching methodology was checked for consistency in the continuous data case and deviations in outcomes versus the ``soft'' approach were found to be minimal.
The method is also robust in the presence of empty clusters, though this issue is not encountered for any of the motivating data sets considered. While it is true that no method for correcting for label switching is guaranteed to provide an optimal solution under all circumstances, the results presented demonstrate that the ``soft'' approach performs well for the real and simulated data considered.

When candidate starting values for $\mathbf{Z}$ are proposed by models with a larger dimension, additional steps are required before the candidate models can be averaged. In this case, the mapping between soft cluster assignments is not one-to-one, but many-to-one. Multiple groups that map to a single group of the reference $\mathbf{Z}$ matrix are merged together, and the values are then averaged in the usual manner. The use of such an approach was investigated, but did not find an example where the performance of BIA improved by including such candidate models for any of the motivating data sets presented.

A further difficulty arises when the number of groups $G$ is large, for example, if $G=10$ or larger. In this case the number of permutations required in order to obtain an exact matching solution quickly becomes infeasible. One computationally cheaper alternative to the exact matching solution is to use a greedy search method that is based on random permutation of the group labels. Multiple permutations can be tried and the best solution retained. Another alternative is to use the \emph{rowmax} method, whereby each row of the cross-classification table is independently mapped to its corresponding column with maximum agreement. Both approaches have drawbacks in comparison to exact matching: the greedy search may obtain a suboptimal matching solution; while perhaps more seriously the \emph{rowmax} method does not guarantee that the one-to-one correspondence between clustering solutions is preserved.

While this issue of class matching for large numbers of groups does not apply for any of the motivating data sets presented, the performance of BIA on the Simulated Data $2$ data set ($G=4$) was investigated, using both greedy search (with only three random starts) and \emph{rowmax} methods. In both cases, the results were unchanged from those reported in Section~\ref{section4_sub_sim2} when exact matching was used.

\section{Results}
\label{section4_results}
The results in this section were obtained using a varying number of initial $\mathbf{Z}$ matrices and preliminary EM runs for each data set and both label switching methods. Each individual initial $\mathbf{Z}$ matrix was obtained by randomly assigning each observation in the data to one of the available groups, with the process  repeated until the required number of $\mathbf{Z}$ matrices was obtained. The BIA code was first run for a range of different values of the number of initial $\mathbf{Z}$ matrices and the number of preliminary EM runs and the optimum combination was then used to produce the results below. For some data sets, this number of initial $\mathbf{Z}$ matrices and preliminary EM steps may seem computationally costly. However, superior BIA performance can still be witnessed over competing methods at a much lower computational cost by reducing the number of $\mathbf{Z}$s and preliminary EM steps, but not as frequently. All $\mathbf{Z}$ matrices are run until full convergence of the EM algorithm. The early iterations of any EM algorithm are usually the most profitable in terms of driving the likelihood function uphill. The BIA method exploits this since it focuses computational power on these early iterations prior to weighting the processed $\mathbf{Z}$ matrices to form a single $\mathbf{Z}$ for use in the EM algorithm.

\subsection{Australian Institute of Sports (AIS) data}
\label{section4 sub AIS}

\begin{figure}[!h]
\begin{center}
\begin{tabular}{cccc}
\subfigure[]{\label{fig:ais_hist_old}\includegraphics[width=5cm, height=5cm]{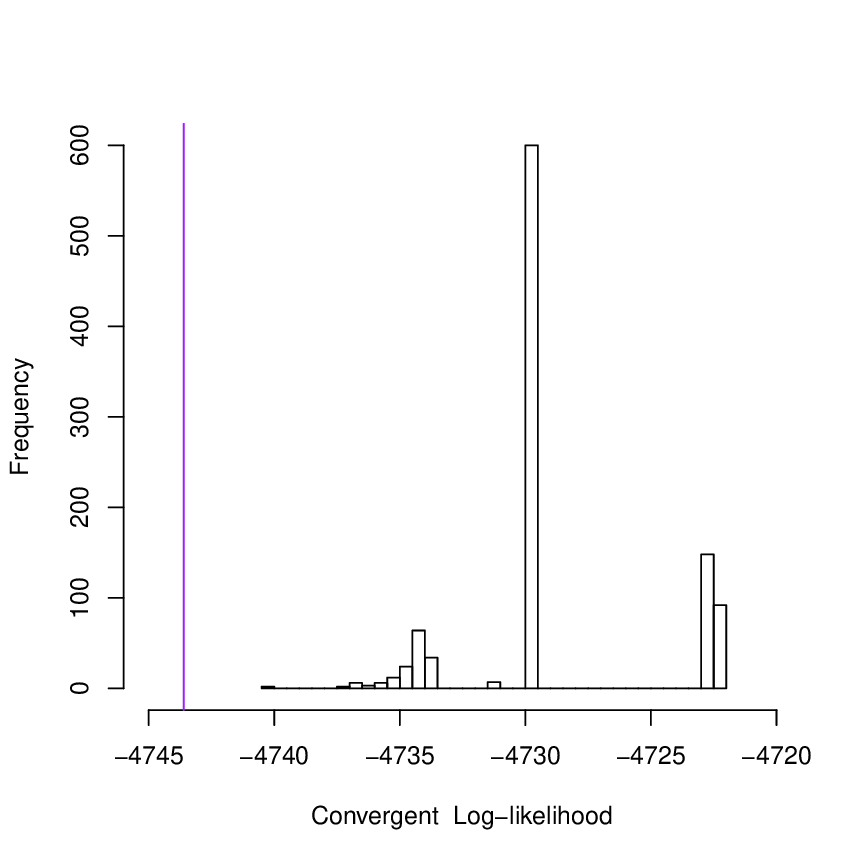}}&
\subfigure[]{\label{fig:ais_hist_new1}\includegraphics[width=5cm, height=5cm]{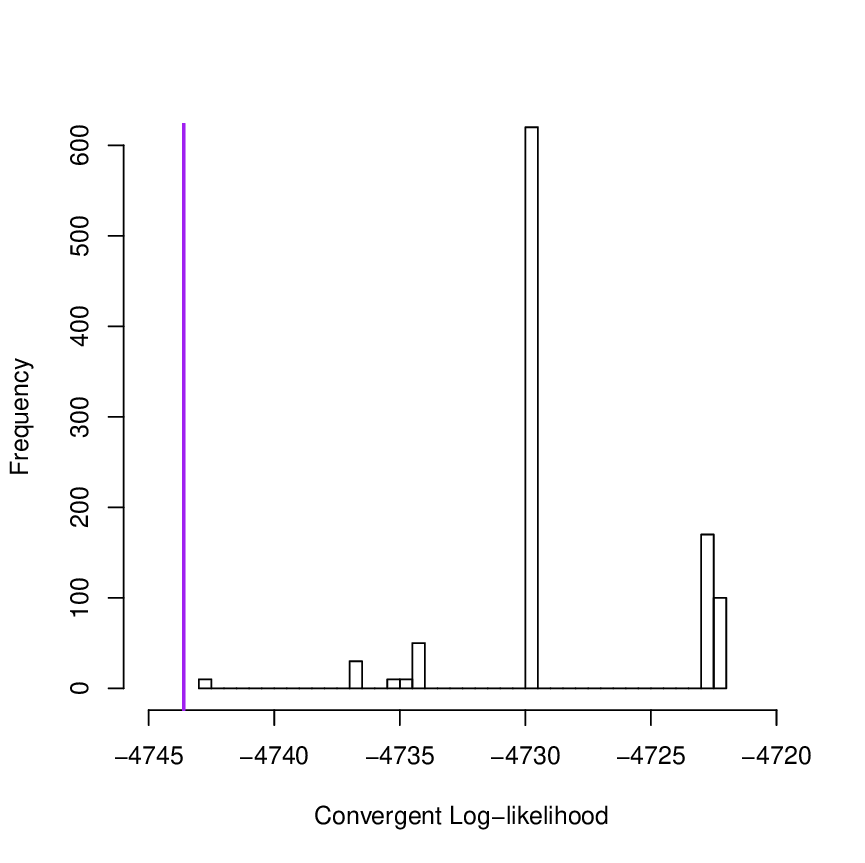}}\\
\subfigure[]{\label{fig:ais_hist_new2}\includegraphics[width=5cm, height=5cm]{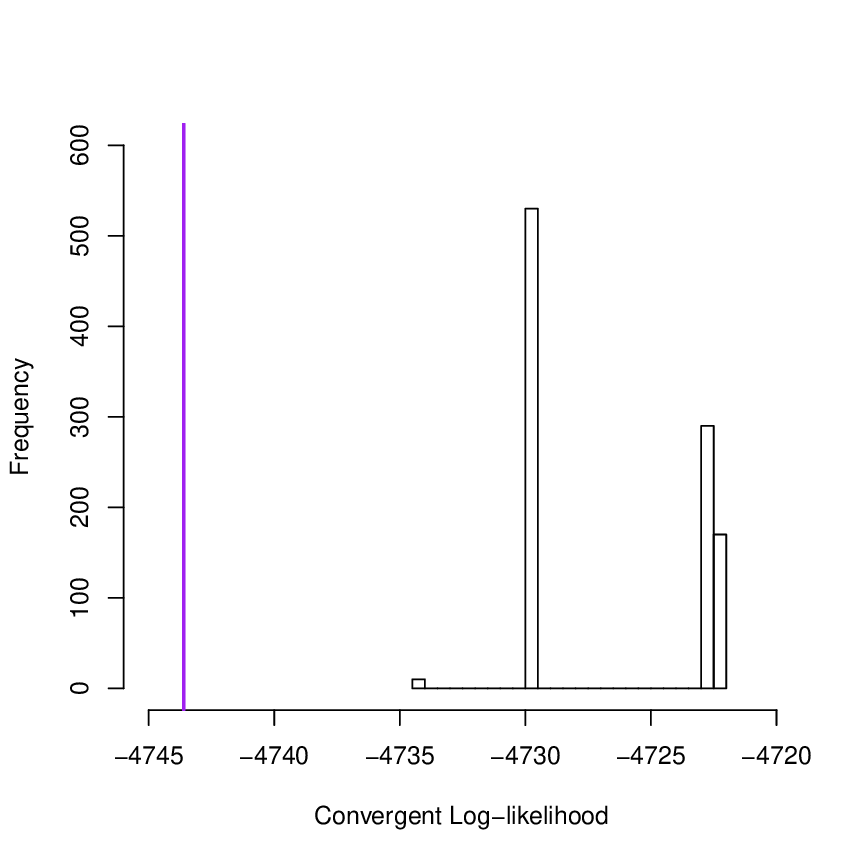}}&
\subfigure[]{\label{fig:ais_hist_random}\includegraphics[width=5cm, height=5cm]{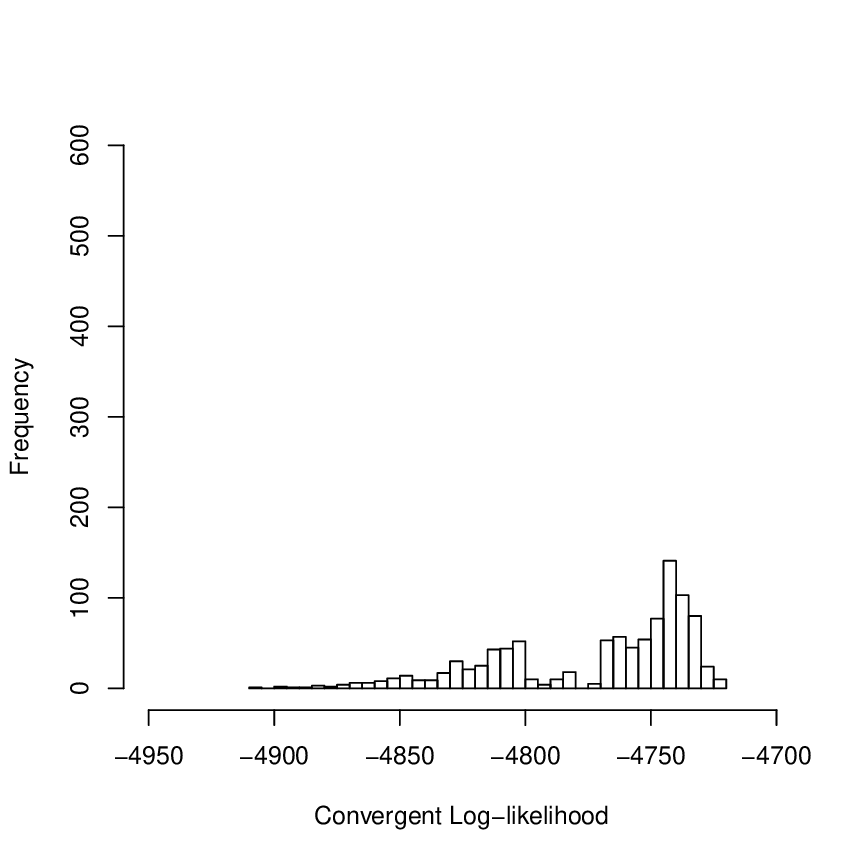}}&
\end{tabular}
\caption{\emph{AIS} data: histogram of convergent log-likelihoods with $G = 2$ and ellipsoidal covariance structure with equal shape and volume (EEV) using (a) pyramid burn-in as detailed in \cite{OHagan2012} (b) BIA with $40$ initial $\mathbf{Z}$ matrices and $50$ preliminary EM runs (c) BIA with $50$ initial $\mathbf{Z}$ matrices and $100$ preliminary EM runs (d) random starting values for the $\mathbf{Z}$ matrices. The purple vertical line at $-4743.6$ represents the optimal log-likelihood reached by \textbf{mclust}'s hierarchical clustering initialization.}
\label{fig:ais_histograms}
\end{center}
\end{figure}

Figure $\ref{fig:ais_histograms}$ displays a histogram of the convergent log-likelihoods for the $11$ continuous variables of the \emph{AIS} data set, using BIA, with both $40$ initial $\mathbf{Z}$ matrices and $50$ preliminary EM runs (Figure $\ref{fig:ais_hist_new1}$) and $50$ initial $\mathbf{Z}$ matrices and $100$ preliminary EM runs (Figure $\ref{fig:ais_hist_new2}$) respectively. In order to produce the plot, each approach is applied $1000$ times and the convergent log-likelihood recorded at the end of each run. The former approach gives a similar distribution of convergent log-likelihood values to that obtained using the previously developed pyramid burn-in methodology developed by \citep{OHagan2012} in Figure $\ref{fig:ais_hist_old}$. The latter gives a superior distribution of convergent log-likelihoods. All three plots contrast strongly with that produced using random starting values for the $\mathbf{Z}$ matrices, where the dominant mode is rarely attained and vastly inferior modes are much more common (Figure $\ref{fig:ais_hist_random}$). The BIA approach was also compared to the simulated annealing method. When applied $1000$ times using highly conservative parameters $\nu^0=0.001, r=0.99, \mbox{ and } s=100,$ in all cases the method converged to an inferior mode of $-4729.5.$

A grid search across possible combinations identified that any combination incorporating upwards of $50$ initial $\mathbf{Z}$ matrices and $100$ preliminary EM runs yields a superior distribution of convergent log-likelihoods for the BIA method in comparison to those identified under competing methodologies. The combination of $40$ $\mathbf{Z}$ matrices and $50$ preliminary EM steps is included merely to demonstrate that even at lower settings the outcome is not inferior to that of competing methodologies.

The model applied has $G = 2$ multivariate normal components and ellipsoidal covariance structure with equal shape and volume (EEV in standard \textbf{mclust} nomenclature) as selected by BIC. The purple vertical line represents the optimal log-likelihood reached by \textbf{mclust}'s hierarchical clustering initialization with $G = 2$ groups and EEV covariance structure. The ``soft'' label-switching methodology is employed in this instance, but there is minimal effect on the results of alternatively using the ``hard'' label-switching approach.

The BIA approach is extremely fast to run, requiring only $5\%$ of the processing time needed by the alternative schemes developed by \cite{OHagan2012} for the \emph{AIS} data. This shows that BIA has potential to be applied to data sets with large numbers of observations and variables, or to more complex mixture models for continuous data. For thoroughness, the BIA method was also applied to the \emph{Virginicas}, \emph{Galaxies} and \emph{Hidalgo} data sets used as motivating examples by \cite{OHagan2012}. As was the case with the \emph{AIS} data, the distribution of convergent log-likelihoods was comparable to or better than that found using the competing burn-in and parameter targeting methods, but at considerably lower computational cost.

\begin{figure}[tbp]
\begin{center}
\includegraphics[scale=0.8]{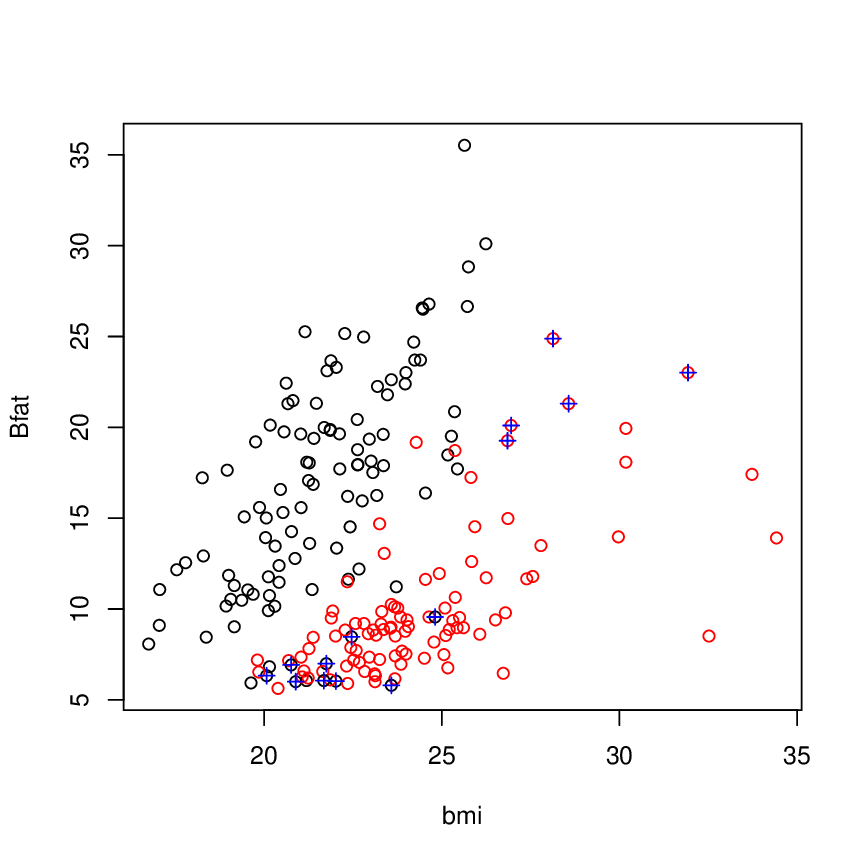}
\caption{\emph{AIS} data: Pairs plot for body mass index (bmi) and body fat Percentage (Bfat) showing the clustering solution using \textbf{mclust}'s hierarchical clustering initialization. The $14$ observations that change cluster membership when using the optimal likelihood identified by BIA are overlaid with crosses.}
\label{fig:ais_clustering}
\end{center}
\end{figure}

Figure $\ref{fig:ais_clustering}$ displays the pairs plot for body mass index (bmi) and body fat Percentage (Bfat) under the clustering solution obtained using a hierarchical initialization of the EM algorithm (convergent log-likelihood of $-4,743.6$). Compared to the optimal convergent log-likelihood that is regularly identified by the BIA method ($-4,722.4$), $14$ data points change cluster membership, highlighted as crosses. The change in clustering solution produces a shift in the total within-cluster sum of squared distances to the cluster mean from $619,110$ for the convergent log-likelihood identified by hierarchical clustering initialization to $606,377$ for the global mode identified by BIA. This example shows that converging to a higher log-likelihood can change the optimal clustering solution of a data set and provide a more intuitive result with clearer separation between groups. Increasingly higher log-likelihoods at convergence may be associated with increasingly more marked divergences in clustering solutions, which at a very minimum should be evaluated for the insights they offer. This demonstrates the importance of converging to the highest possible meaningful log-likelihood and the inherent value of the BIA method.

\subsection{Simulated Data $1$: mixture of Gaussians}
\label{section4 sub simdata}
When a multivariate mixture of Gaussians model with the most general covariance structure is fitted to Simulated Data $1$ as per \cite{Baudry2015}, hierarchical clustering initialization identifies a $3$ group solution as optimal in terms of BIC for $98$ of the $100$ data sets tested. Random starts gives a similar outcome, favouring a $3$ group solution for $95$ of the $100$ cases. The same mixture of Gaussian distributions initialized using BIA identifies a $3$ group solution as optimal in terms of BIC in $49$ of the $100$ cases and a $2$ group solution as optimal in the remaining $51$ cases. This is preferable in such a case where both the $2$ group and $3$ group solutions are viable, demonstrating to the user that the underlying horizontal and vertical data structure can be recreated with one less component if so desired. Hence BIA can be seen to offer similar benefits to the Supervised Integrated Complete Likelihood (SICL) decision criterion introduced in \cite{Baudry2015}. However, whereas \cite{Baudry2015} use the SICL rather than the BIC to form the model decision criterion, under the BIA approach BIC is retained as the model decision criterion. Instead, BIA reaches updated decisions as to the nature of the selected model via superior model initialization. Use of pyramid generated starting values produces an outcome that is between that of hierarchical clustering/random starts and BIA, identifying a $3$ group solution as being best in $81$ of the $100$ cases.

Minimal variation from the trend observed for $n = 200$ is seen even as the sample size is increased for Simulated Data $1$. BIA continues to substantially outperform competing methods in terms of ability to identify the $2$ group solution. Table $\ref{tab:simstudy1_results}$ reports the number of cases across the $100$ simulated data sets that the $G=2$ solution is identified for each of the sample sizes $n = 200, 500, 1000$ and $5000$.

\begin{table}[]
\caption{Number of times, out of 100, that $G = 2$ is obtained using random starts, deterministic annealing, and BIA across different sample sizes for Simulated Data $1$.}
\centering
\begin{tabular}{|l|l|l|l|l|}
\hline
Initialization method   & n = 200 & n = 500 & n = 1000 & n = 5000 \\ \hline
Random Starts           & 5       & 7       & 6        & 6        \\ \hline
Hierarchical Clustering & 2       & 2       & 3        & 1        \\ \hline
Pyramid Burn-in         & 19      & 21      & 23       & 22       \\ \hline
BIA                     & 51      & 50      & 48       & 49       \\ \hline
\end{tabular}
\label{tab:simstudy1_results}
\end{table}

\subsection{Simulated Data $2$: LCA data}
\label{section4_sub_sim2}

Firstly, results for sample size $n = 500$ are reported. Table~\ref{tab:simstudy2_results} shows the different solutions obtained across $100$ initializations arising from each of the competing methods. All methods recover the dominant mode more regularly than simple random starts ($51$\% of the time), with BIA ($100$\% recovery) performing more favourably than deterministic annealing ($89$\%).

The results for BIA were obtained by using $10$ random starts and $10$ iterations of the EM algorithm before BIA was applied. For the annealing method, the tuning parameters were set to be $\nu^0=0.05$,  $r=0.96$ and $s=10$. In this case, it is found that the annealing method was extremely sensitive to the specification of $r$: if $r$ is set to $0.95$ or $0.94$, with the other parameters kept the same, the dominant mode is uncovered less frequently, $79$\% and $72$\% of the time respectively (still an improvement over using simple random starts). Similar results were obtained using BIA when fewer starting values or fewer initial iterations were used. The pyramid burn in approach performs almost as well as BIA ($98\%$ recovery) but takes $4$ times longer to achieve this result.

\begin{table}[ht]
\caption{Distribution of convergent log-likelihoods for Simulated Data $2$ with $n = 500$ using $100$ initializations based on random starts, deterministic annealing, BIA and pyramid burn-in.}
\centering
\begin{tabular}{lrrrrrrr}
\hline
Log-likelihood & -4729 & -4728 & -4727 & -4726 & -4725 & -4724 & -4723 \\
\hline
Random Starts & 3 & 1 & 10 & 8 & 0 & 27 &  51 \\
Annealing & 0 & 0 & 9 	& 0 & 0 & 2 & 89 \\
Pyramid & 0 & 0 & 0 & 2 & 0 & 0 & 98 \\
BIA &   0 & 0 & 0 & 0 & 0 & 0 & 100 \\
\hline
\end{tabular}
\label{tab:simstudy2_results}
\end{table}

Starting values based on hierarchical clustering solutions using a binary dissimilarity metric and the eight different linkage methods provided by the \textbf{hclust} function in \textbf R were also investigated. Of these, the dominant mode was obtained only when Ward's linkage method was used. Five of the eight linkage methods obtained worse solutions than any of those obtained by using simple random starts.

\begin{figure}[!h]
\begin{center}
\begin{tabular}{cc}
\subfigure[]{\label{fig:simstudy2_fits_inferior}\includegraphics[scale=0.41]{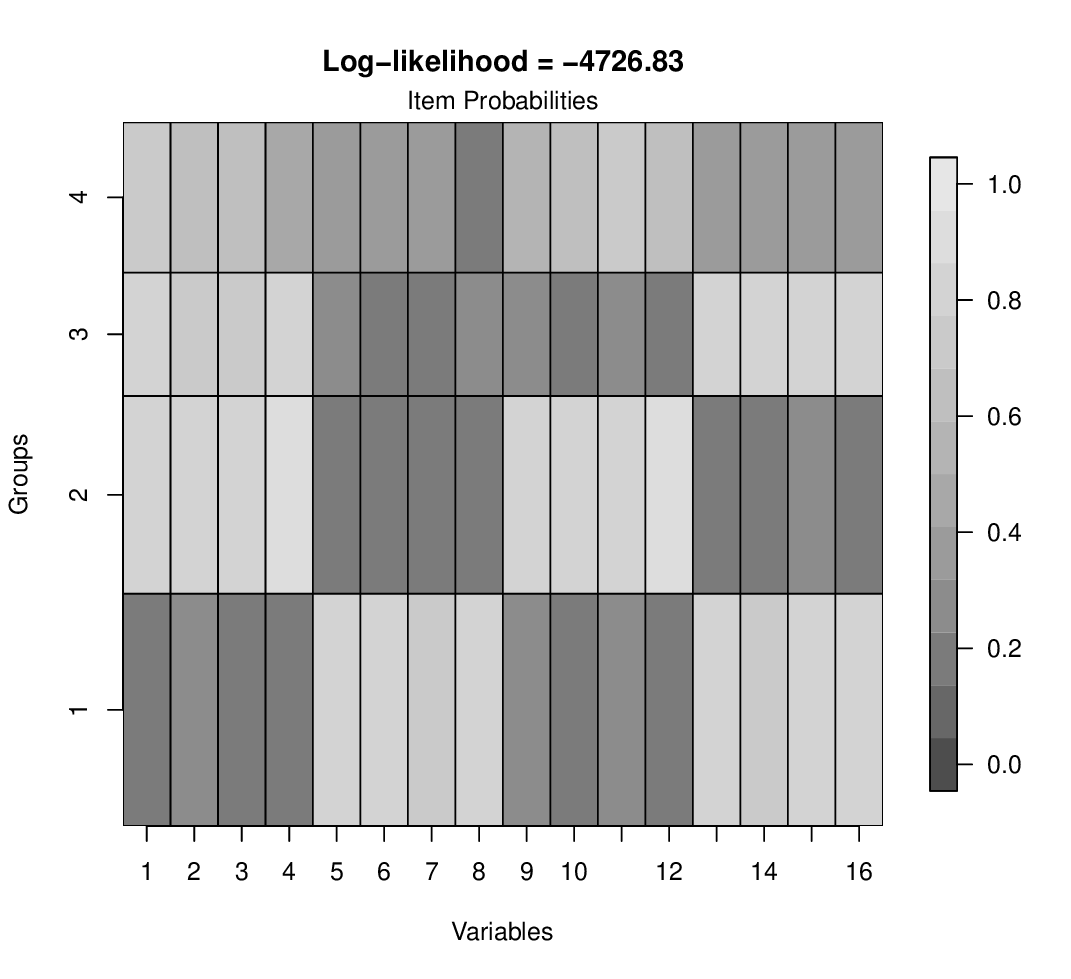}}&
\subfigure[]{\label{fig:simstudy2_fits_dominant}\includegraphics[scale=0.41]{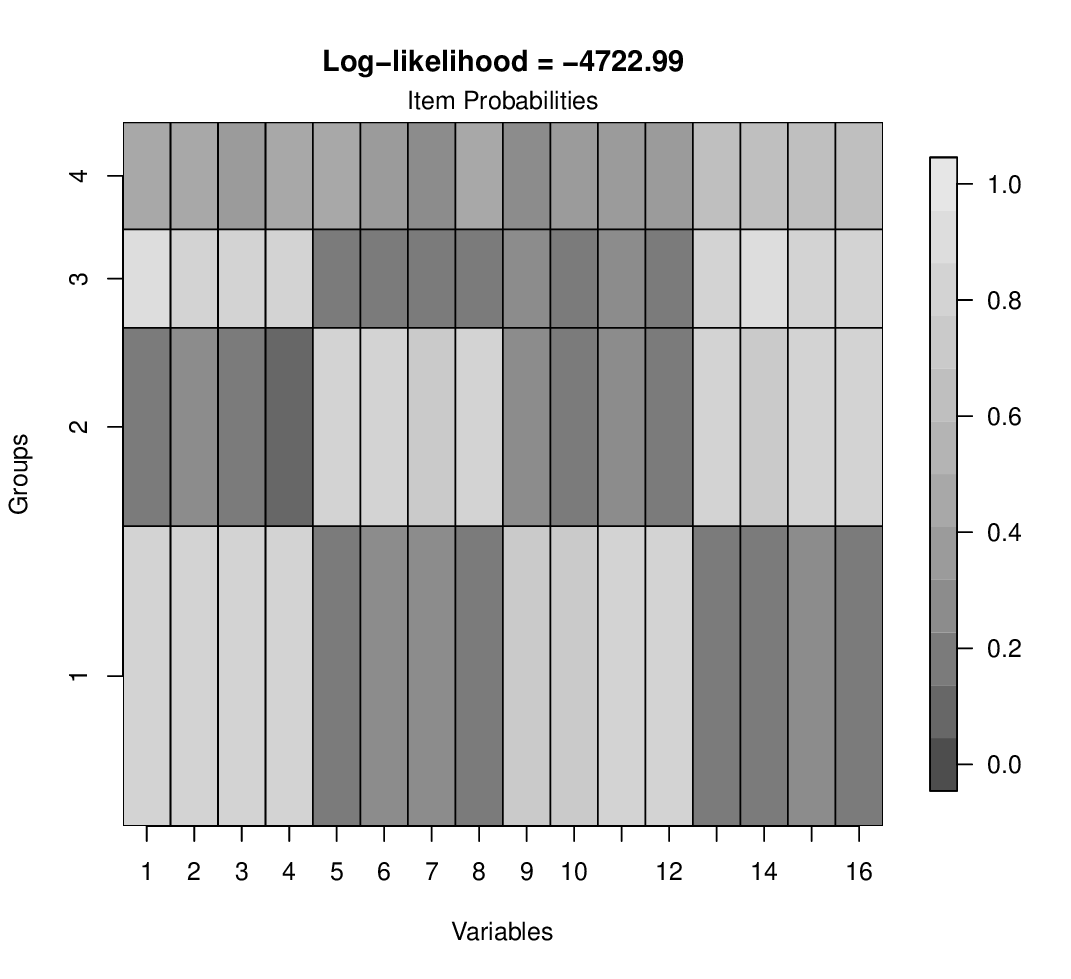}}\\
\end{tabular}
\caption{Visualisation of different parameter estimates obtained by local maxima for Simulated Data $2$ at (a) an inferior mode and (b) the dominant mode.}
\label{fig:simstudy2_fits}
\end{center}
\end{figure}

Competing fits to the data are visualised in Figure~\ref{fig:simstudy2_fits}. The solution obtained at the dominant mode (Figure $\ref{fig:simstudy2_fits_dominant}$) is clearly similar to the true model structure visualised in Figure~\ref{fig:simstudy2}. In contrast the fitted parameters obtained at an inferior mode (Figure $\ref{fig:simstudy2_fits_inferior}$) would suggest a different interpretation of the data, especially for Groups $3$ and $4$. There is also a considerable amount of disagreement ($21$\%) between the clustering solutions obtained at the different modes, highlighting the value of BIA's ability to routinely identify the dominant mode.

Results for different sample sizes are shown in Table~\ref{tab:simstudy2b_results}. For smaller sample sizes, the pyramid approach achieves the best results, while BIA also outperforms annealing and random starts. For large sample sizes, all three methods clearly outperform simple random starts. For $n = 100,$ the performance for BIA and pyramid methods are superior to simple random starts and annealing methods, with the dominant mode obtained $53\%$ and $58\%$ of the time compared to $17\%$ and $14\%$ respectively. For $n=250$, the performance of the pyramid method is best, with the dominant mode obtained $92\%$ of the time, compared to $74\%$ for BIA, $41\%$ for annealing and $38\%$ for random starts. For the largest sample sizes, the random starts performance is poor, recovering the dominant mode 10\% of the time for $n = 1000$, and in no cases when $n = 5000$. For these sample sizes, pyramid, annealing and BIA methods recover the dominant mode in all cases. Across the sample sizes, the dominant mode was obtained using hierarchical clustering with Ward's linkage methods for all sample sizes except $n=100$. For all sample sizes, several linkage methods obtained highly inferior local modes.

\begin{table}[ht]

\caption{Number of times, out of 100, that dominant modes were obtained using random starts, deterministic annealing, and BIA across different sample sizes for Simulated Data $2$ and Simulated Data $3$.}
\centering
\begin{tabular}{lrrrr|rr}
& Simulated & \,\,\,Data & $2$ & & Simulated & Data $\,\,\,\,\,\,\,3$ \\
  \hline
$n = $ & 100 & 250 & 1000 & 5000 & 1000 & 5000 \\
  \hline
Random Starts &  14 & 38  &  10 & 0 & 12 & 0\\
Annealing &  34 &  41 & 100 & 100 & 0 & 0\\
Pyramid & 58  &  92 & 100 & 100 & 32 & 0 \\
BIA &  52 &  74 & 100 & 100 & 71 & 55\\
   \hline
\end{tabular}

\label{tab:simstudy2b_results}
\end{table}

For the imbalanced simulation study (Simulated Data $3$), the BIA performance across the $100$ initializations tested is superior to that of the competing methods. For $n = 1000$, the dominant mode is recovered in 12\% of cases using simple random starts, 32\% of cases using the pyramid method, 71\% of cases using BIA with $50$ random starts run for $50$ iterations, and in no cases by any other method. For $n=5000$, the dominant mode is obtained by BIA using $50$ random starts run for $10$ iterations in 55\% of cases. No other method recovers the dominant mode for this sample size.

\subsection{Carcinoma data}
\label{section4 sub Carcinoma}

\begin{figure}[tbp]
\begin{center}
\includegraphics[scale=0.5]{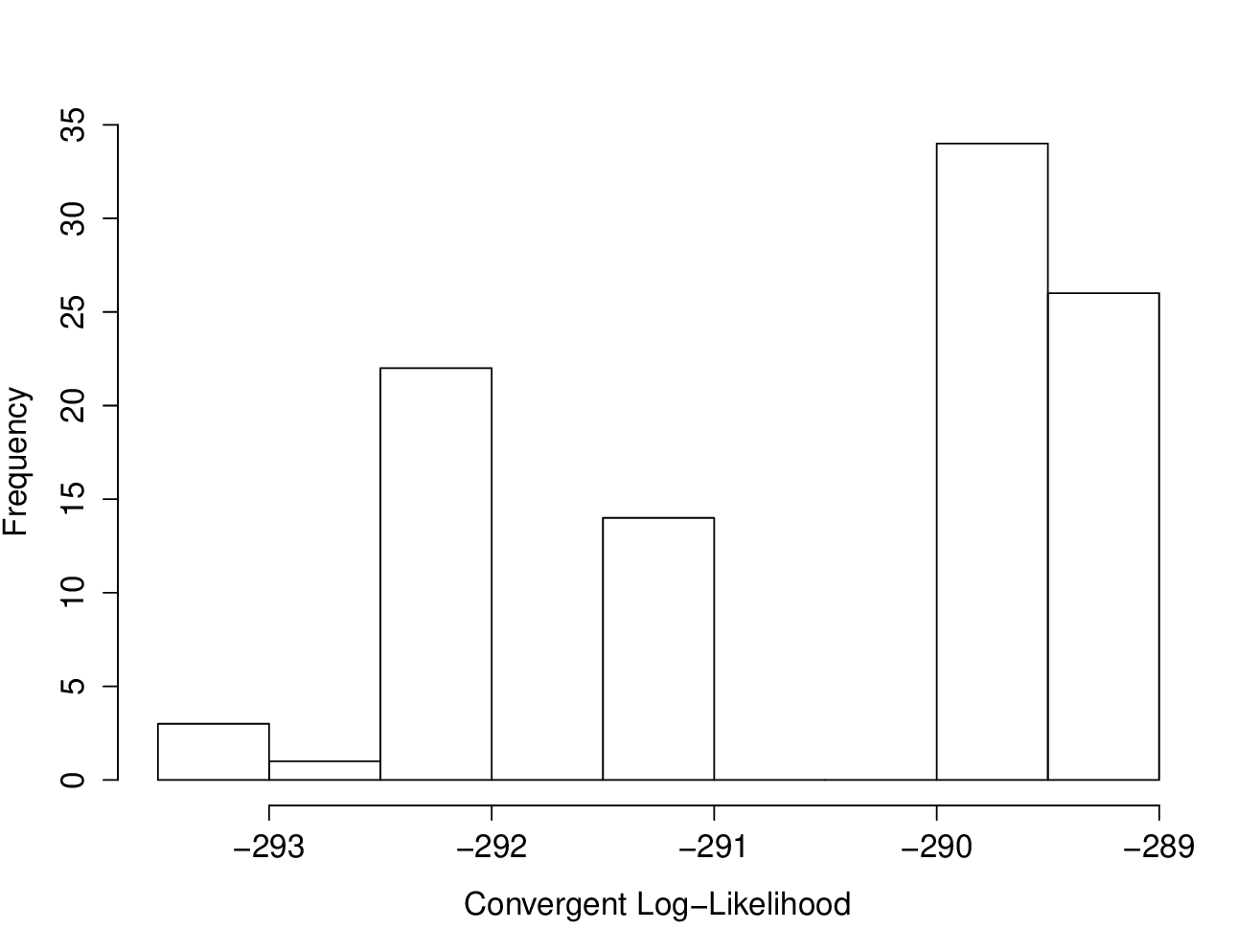}
\end{center}
\caption{Histogram of distribution of convergent log-likelihoods using $100$ random starts for the \emph{Carcinoma} data.}
\label{fig:carc_hist}
\end{figure}

\begin{figure}[tbp]
\begin{center}
\includegraphics[scale=0.5]{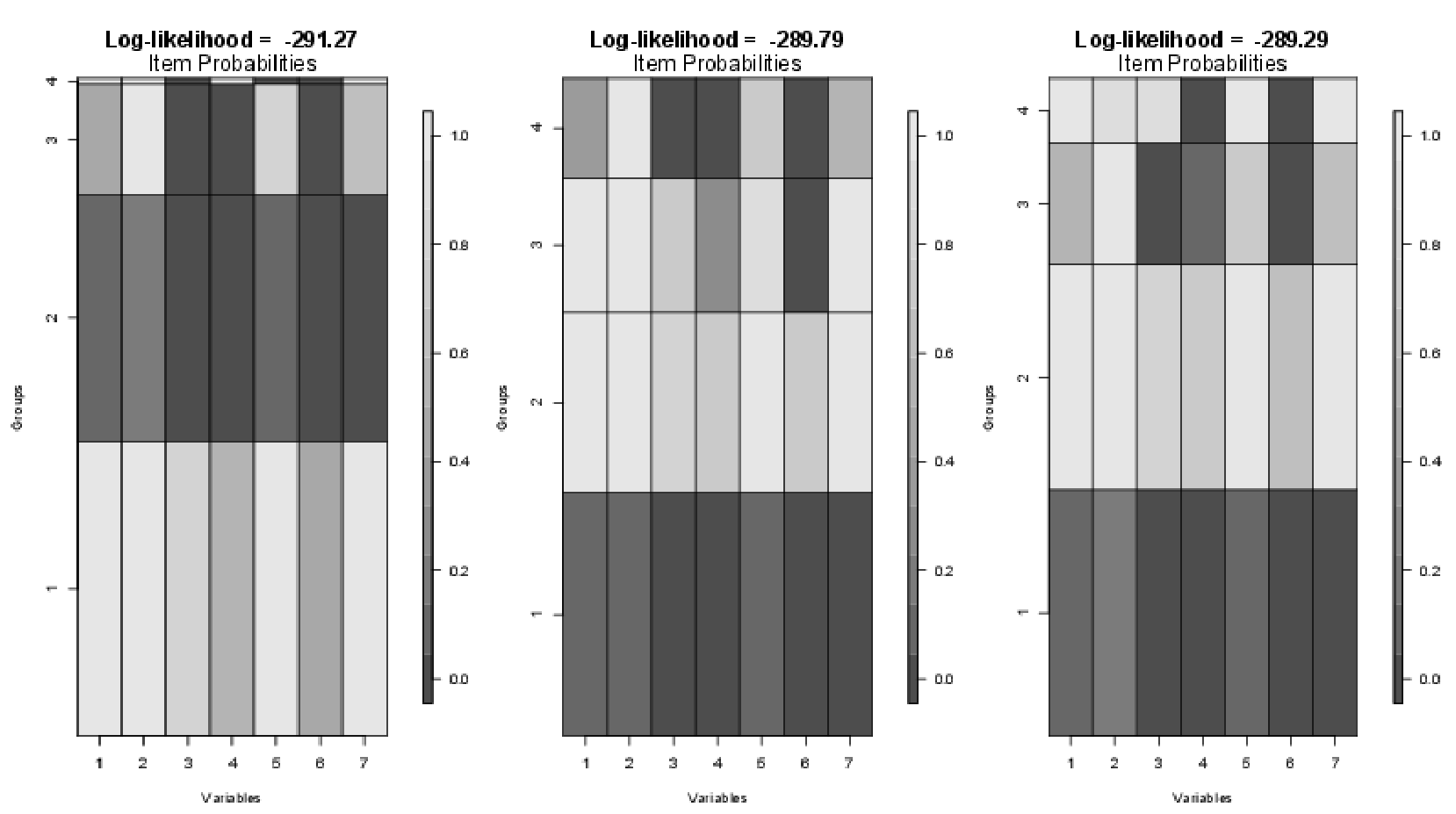}
\end{center}
\caption{Visualisation of different parameter estimates obtained by local maxima for the LCA model applied to the \emph{Carcinoma} data.}
\label{fig:carc_fits1}
\end{figure}

For each of $100$ randomly generated starting values, the EM algorithm was run to convergence for the \emph{Carcinoma} data and multiple local maxima were uncovered. The distribution of log-likelihoods is shown in Figure $\ref{fig:carc_hist}$. Parameter estimates obtained by the LCA model are visualised in Figure $\ref{fig:carc_fits1}$. 

As can be seen in Figure $\ref{fig:carc_fits1}$, the parameter estimates corresponding to convergent log-likelihood $-289.79$ and convergent log-likelihood $-289.29$ are highly similar, 
however, a different interpretation of the data corresponds to convergent log-likelihood $-291.27$. This illustrates the importance of not converging to the sub-optimal mode.

As seen in Table $\ref{tab:carcinoma_results}$, BIA achieves the global maximum convergent likelihood 92\% of the time, and the second best, highly similar mode the rest of the time, using $40$ random starts and $20$ iterations of the algorithm before BIA is applied. This is a marked improvement over the outcome when random starts are employed. 

In this case, the performance of BIA is an improvement over using simple random  starts, and comparable, in comparison to the annealing method. Using settings, $\nu^0=0.05$,  $r=0.95$ and $s=10$, \citet{zhou2010} report obtaining the dominant mode in 99\% of cases. Using $32$ candidate starting values, the pyramid method also had similar performance.

\begin{table}[ht]
\caption{Distribution of convergent log-likelihoods for $100$ initializations of the \emph{Carcinoma} data using BIA, random starts, deterministic annealing and pyramid burn-in.}
\centering
\begin{tabular}{rrrrrrr}
\hline
Log-likelihood & -294 & -293 & -292 & -291 & -290 & -289 \\
\hline
Random Starts &   1 &   3 &  21 &  12 &  40 &  23 \\
Annealing &   0 &  1 &  0 &  0 &  0 &  99 \\
BIA &   0 &   0 &  0 &  0 &  8 &  92 \\
Pyramid &   0 &   0 &  0 &  0 & 27 &  73 \\
\hline
\end{tabular}
\label{tab:carcinoma_results}
\end{table}

Using starting values based on hierarchical clustering solutions with a binary dissimilarity metric, only inferior modes were obtained, regardless of the linkage method that was used.

\subsection{Alzheimer's data}
\label{section4 sub Alzheimer's}
\begin{figure}[tbp]
\begin{center}
\includegraphics[scale=0.5]{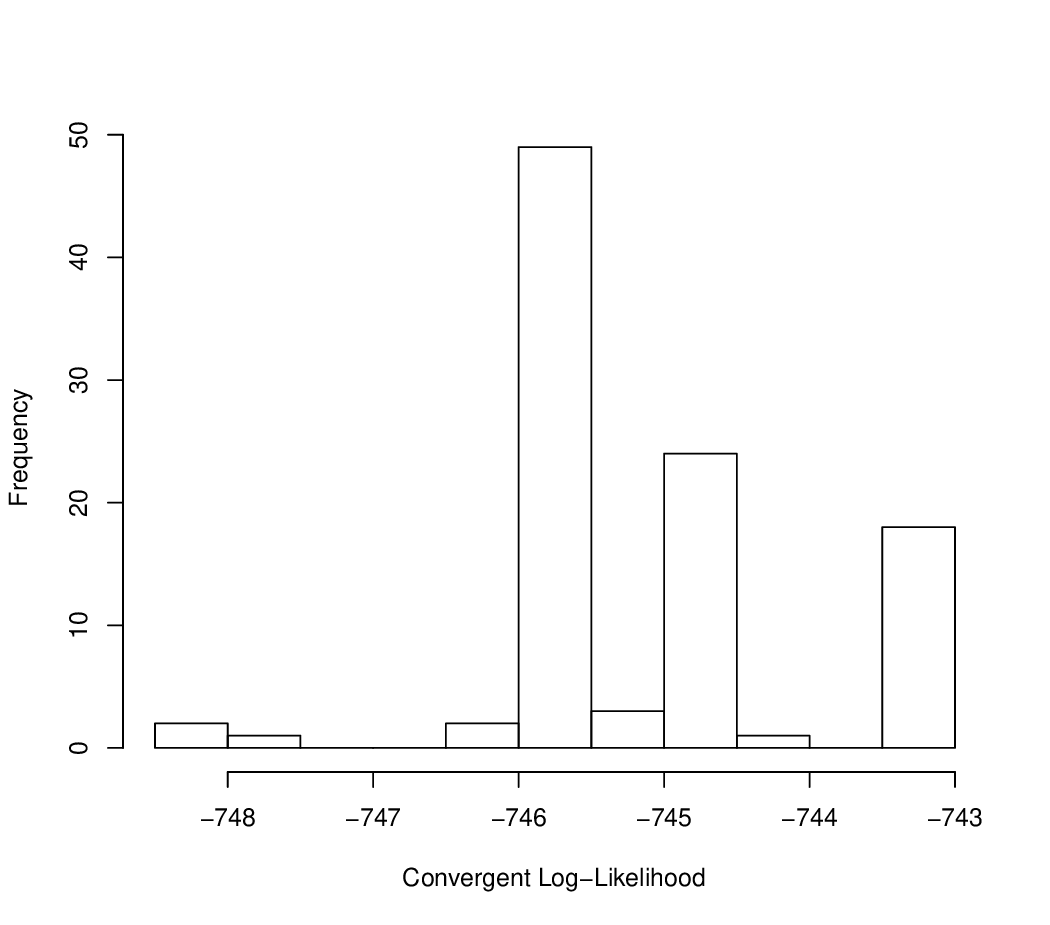}
\end{center}
\caption{Histogram of distribution of convergent log-likelihoods using $100$ random starts for the \emph{Alzheimer's} data.}
\label{fig:alz_hist}
\end{figure}

A $3$ group LCA model was applied to the Alzheimer's data from $100$ randomly generated starting values. The distribution of the convergent log-likelihoods is shown in Figure $\ref{fig:alz_hist}$; the global maximum is only obtained in 18 cases. Table $\ref{tab:Alz_tab1}$ displays the differing solutions arising from applying LCA to the \emph{Alzheimer's} data at convergent log-likelihoods of $-745.7$ and $-743.5$ respectively. Again, the latter solution produces a significant alteration in interpretation of symptom composition in the three different classes, with ramifications for subsequent patient classification.

\begin{table}[ht]
\caption{Parameter estimates for the 3 group LCA model fitted to the \emph{Alzheimer's} data. Firstly, the sub-optimal convergent log-likelihood $-745.7$; and secondly, the global convergent log-likelihood $-743.5$.}
\centering
\begin{tabular}{rrrrrrrr}
\hline
  & $\hat{\tau}_g$ & Hallucination & Activity & Aggression & Agitation & Diurnal & Affective \\
\hline
  1 & 0.39 & 0.13 & 0.70 & 0.28 & 0.00 & 0.30 & 0.78 \\
  2 & 0.31 & 0.03 & 0.46 & 0.00 & 0.18 & 0.04 & 0.51 \\
  3 & 0.30 & 0.07 & 0.80 & 0.41 & 1.00 & 0.38 & 0.97 \\
\hline
  1 & 0.51 & 0.06 & 0.52 & 0.06 & 0.13 & 0.10 & 0.55 \\
  2 & 0.47 & 0.10 & 0.79 & 0.37 & 0.60 & 0.37 & 1.00 \\
  3 & 0.02 & 0.00 & 0.82 & 1.00 & 0.21 & 1.00 & 0.00 \\
\hline
\end{tabular}
\label{tab:Alz_tab1}
\end{table}

The performances of the BIA and annealing methods are compared to that of random starts in Table $\ref{tab:Alz_comparison_tab}$. Each method was run 100 times. In the case of BIA, $20$ random starts run for $200$ iterations were averaged. In this instance, the approach was computationally slower, a single run of the BIA algorithm run taking about 80\% longer than separately running a single run to convergence $100$ times.

For the annealing method, after some experimentation, the tuning parameters were set to be $\nu^0=0.12$,  $r=0.87$ and $s=10$.
While the BIA approach produces an improved performance, with the dominant mode being achieved in $98$\% of cases, the annealing method converges to this maximum only once. Despite an extensive search of the annealing tuning parameter space, it proved impossible to substantively improve this performance. Using $32$ candidate starting values, the pyramid method also struggled to obtain the dominant mode, doing so in only $7$\% of cases.

\begin{table}[ht]
\caption{Distribution of convergent log-likelihoods, rounded to the nearest integer, for $100$ initialisations of the \emph{Alzheimer's} data using random starts, annealing, BIA and Pyramid burn-in.}
\centering
\begin{tabular}{lrrrrr}
\hline
Log-likelihood & -748 & -746 & -745 & -744 & -743 \\
\hline
Random Starts & 3 & 51 &  27 & 1 &  18 \\
Annealing & 0 &  97 &   2 & 0 &   1 \\
BIA &  0 & 0 &  2 & 0 & 98\\
Pyramid &  0 & 93 & 0 & 0 & 7 \\
\hline
\end{tabular}
\label{tab:Alz_comparison_tab}
\end{table}

\begin{figure}[tbp]
\begin{center}
\includegraphics[scale=0.55]{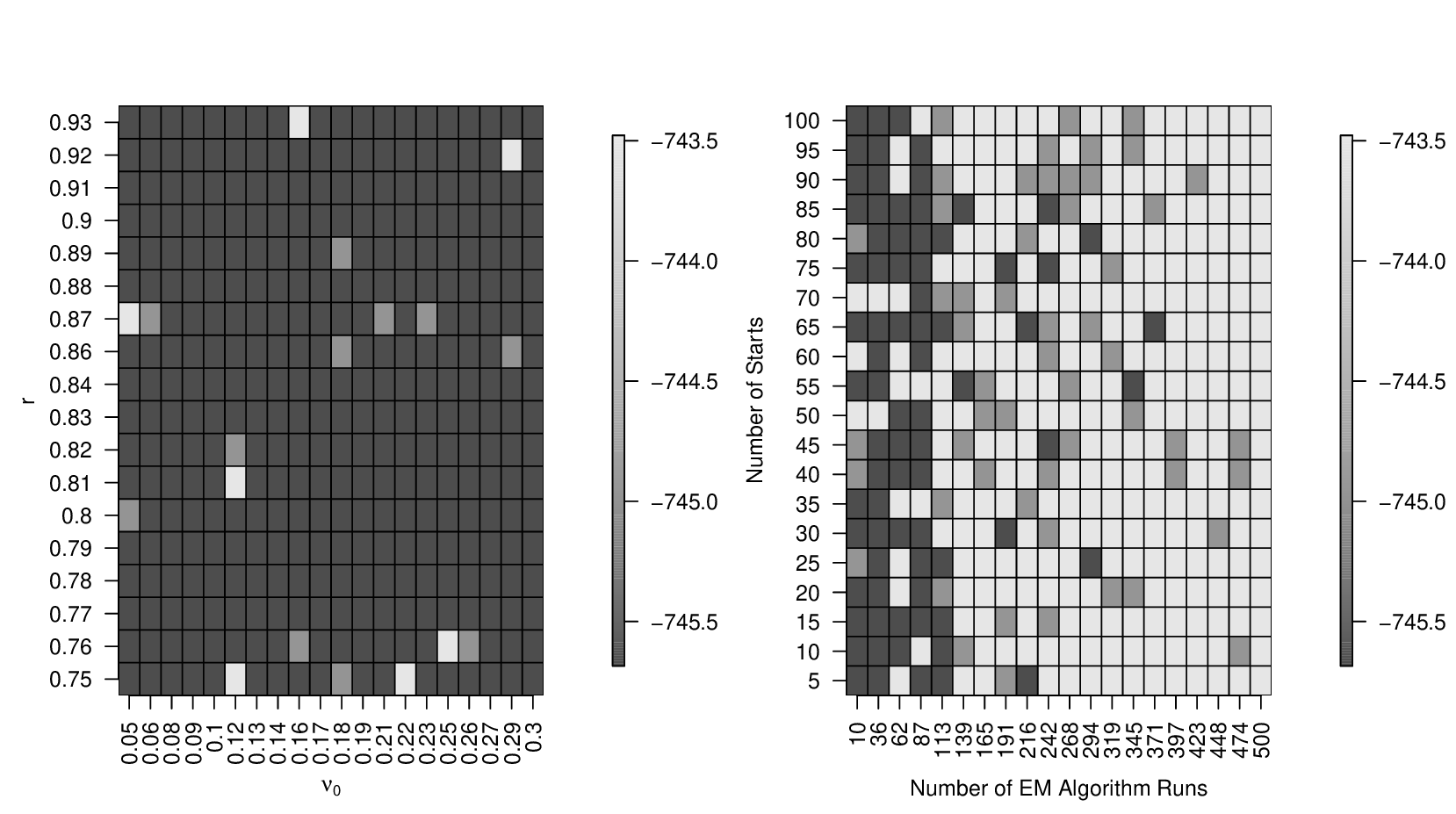}
\end{center}
\caption{Heat map of convergent log-likelihoods obtained using (left) deterministic annealing with different tuning parameter settings and (right) BIA with different tuning parameter settings, for the \emph{Alzheimer's} data. A light grey colour indicates an optimal value of the convergent log-likelihood has been obtained.}
\label{fig:alz_bia_space}
\end{figure}

This is demonstrated in Figure $\ref{fig:alz_bia_space}$, which shows heat maps of the convergent log-likelihoods obtained, firstly using deterministic annealing for different settings of $\nu^0$ and $r$ and secondly using BIA for different numbers of random starts and EM algorithm iterations per random start. Despite using $400$ different parameter combinations, the global mode is rarely discovered using deterministic annealing.\footnote{Using $r = 0.94\mbox{ or } 0.95$ regularly resulted in a saddle point with the value of the log-likelihood converging to -770. These results have been omitted from Figure $\ref{fig:alz_bia_space}$.} Conversely, once the BIA method is allowed to initially run for more than $110$ iterations, the global mode is found consistently, suggesting that the clustering performance of BIA is not as sensitive to the choice of initialization parameters. 

A possible explanation for the poor performance of the annealing method is that in the case of the \emph{Alzheimer's} data the suboptimal solution consists of roughly equal sized groups; the opposite is true in the \emph{Carcinoma} example, where the method's performance is superior to that of BIA. It appears that, unless its tuning parameters are very precisely specified, the annealing method may be biased towards solutions where cluster sizes do not vary significantly.
The propensity for the competing methodologies to attain the global maximum convergent log-likelihood is tested across $100$ matching random starting positions, with the results presented in Table $\ref{tab:Alz_comparison_tab}$. The BIA approach outperforms the basic random starts and annealing methods.

Similarly to Section~\ref{section4 sub Carcinoma}, only inferior modes were obtained when using starting values based on a hierarchical clustering solution with a binary dissimilarity metric, regardless of the linkage method used.

\subsection{Karate data}
\label{section4 sub Karate}
\begin{figure}[tbp]
\begin{center}
\includegraphics[scale=0.5]{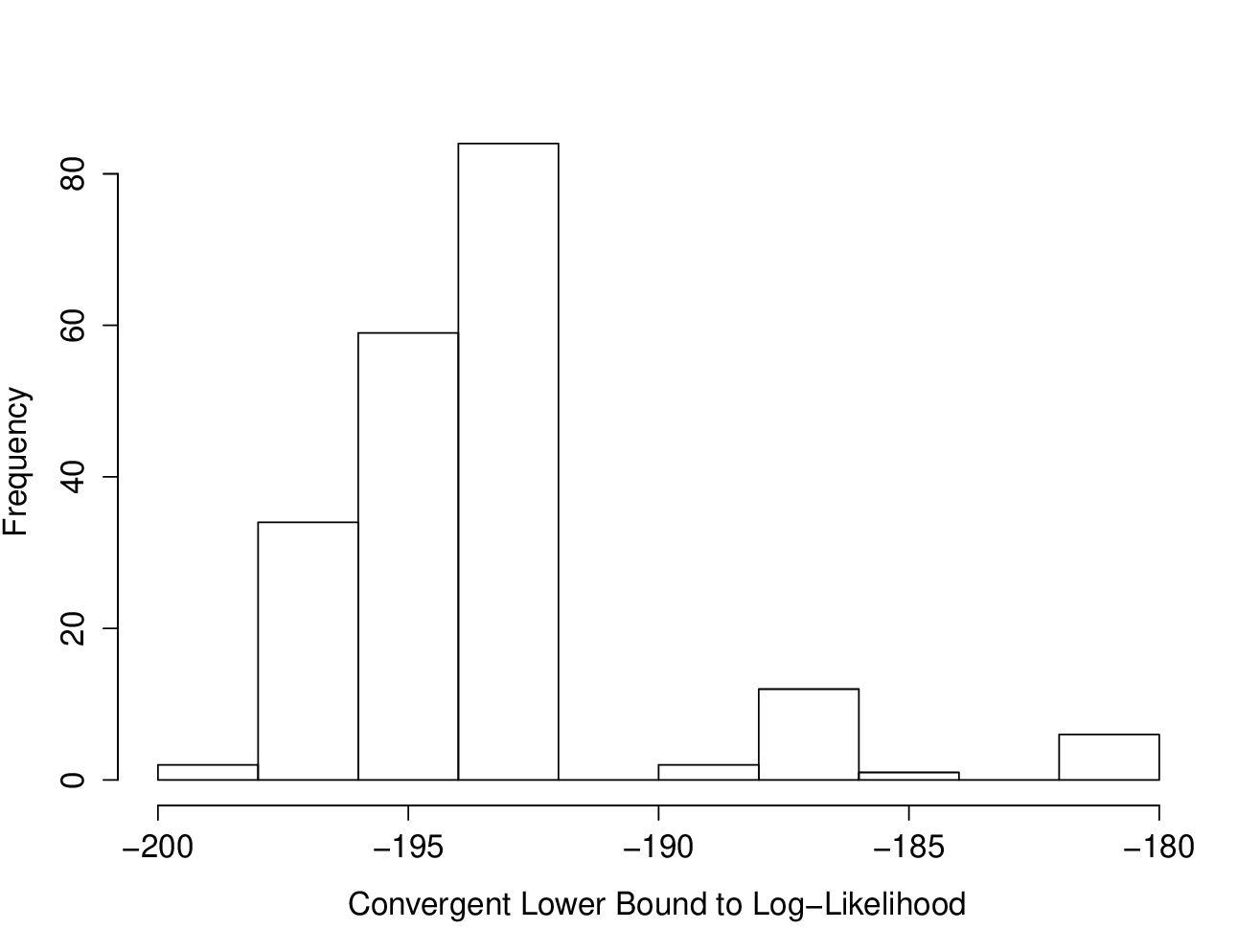}
\caption{Histogram of local maxima obtained for the stochastic blockmodel (SBM) applied to the \emph{Karate} data from $200$ random starting values.}
\label{fig:KarateHist}
\end{center}
\end{figure}

Obtaining the global maximum for SBM applied to the \emph{Karate} data set when $G =4$ using random starts proves to be challenging; over the course of $200$ different runs, the maximum was reached only $4$ times. A histogram of the various local maxima that can be obtained is shown in Figure $\ref{fig:KarateHist}$. Conversely, the BIA approach, using $200$ starting values run initially for $15$ iterations, obtains the global maximum $19$ times out of $20$. These results are shown in Table $\ref{tab:Karate}$. The BIA approach is also much more computationally efficient in this instance; a complete pass of the BIA algorithm is approximately $30$ times faster than the process of having a single starting position run to convergence $200$ times.

\begin{table}[ht]
\caption{Distribution of convergent lower bounds to the log-likelihood {$\mathcal L$} for the Karate data. }
\centering
\begin{tabular}{lrrrrrrrrr}
\hline
$\hat{\mathcal L}$ & -204 & -197 & -195 & -194 & -193 & -188 & -187 & -184 & -180 \\
\hline
Random Starts &   1  & 34 &  45 &  10 &   83 &  19 &   3 &   1  &  4 \\
BIA  & 0 & 0 & 0 & 0 & 0 & 0  & 1 & 0 & 19  \\
\hline
\end{tabular}
\label{tab:Karate}
\end{table}

\begin{figure}[tb]
\begin{center}
\includegraphics[scale=0.5]{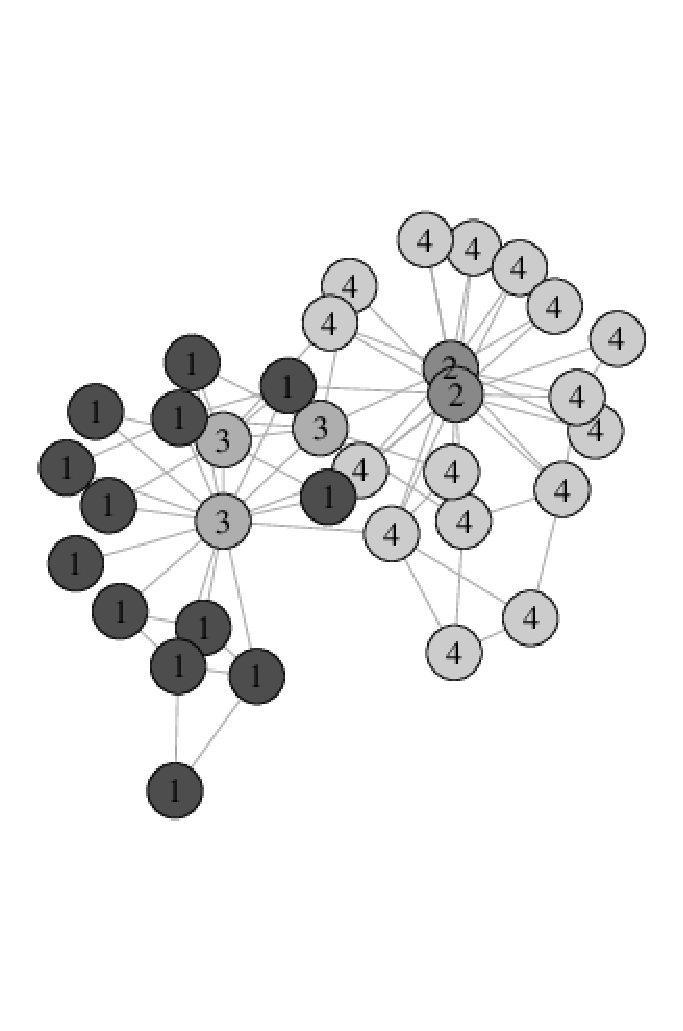}
\caption{Plot of the \emph{Karate} network data set, with $4$ groups fitted to the data.}
\label{fig:Karate4}
\end{center}
\end{figure}

The clustering that obtains the global maximum is visualised in Figure $\ref{fig:Karate4}$. Made clear is the division of the group into two political factions, with each faction having leaders (the nodes labelled 2 and 3, coloured medium-dark and medium-light grey ) and followers (the nodes coloured light and and dark grey, labelled 1 and 4), which mainly connect to their respective leaders. Note that despite being the two largest groups, no links are shared between pairs of nodes labelled 1 and 4. This clustering complements the description of the data in Section $\ref{section2 Karate}$.

\section{Conclusions}
\label{section5_conclusions}

It has been demonstrated that using BIA across a variety of model-based clustering applications can lead to higher convergent likelihoods being reached and that optimal likelihoods can be reached more frequently than under other competing methods. The results from two simulation studies suggest that the method may be especially useful for data sets with large sample sizes. The importance for the clustering solution of reaching the optimal likelihood was also clearly demonstrated. A sub-optimal convergent log-likelihood can lead to an inferior, potentially misleading clustering solution.

The BIA method improves upon \textbf{mclust}'s hierarchical clustering initialization method for the multivariate continuous \emph{AIS} data and reaches a higher convergent likelihood. This higher likelihood in turn changes the clustering solution of the data. While for this example it is not a major change, it does highlight that the clustering solutions obtained by  \textbf{mclust} are not guaranteed to be optimal. BIA also outperforms random starts and pyramid burn-in in terms of the distribution of convergent log-likelihoods attained.

For the Simulated Data $1$ simulation experiment using mixtures of overlapping Gaussian components, BIA correctly identifies both relevant solutions whereas hierarchical clustering initialization and random starts almost always identify only the $3$ group possibility. There is also marked benefit to using BIA versus pyramid burn-in, though the latter is an improvement over random and hierarchical clustering starting values. These outcomes remain consistent as the sample size of the simulated data sets increases.

For the categorical data examples, the BIA results are reasonably close to those of \citet{zhou2010} with respect to the \emph{Carcinoma} data, while the BIA method comprehensively outperforms the pyramid and random starts methods for this data set. For the Simulated Data $2$ simulation experiment BIA is substantially better, slightly better and very marginally better than random starts, annealing and pyramid burn-in respectively. Conversely BIA comprehensively outperforms all competing methods for Simulated Data $3$ and for the \emph{Alzheimer's} data.

For the network data example using the \emph{Karate} data, the BIA approach attains the global maximum likelihood for the $4$-group model far more consistently and efficiently than a simple random starts approach, even when only an approximation to the log-likelihood is available.

While the illustrative data sets used were moderate in terms of size and dimensionality, the BIA approach could prove increasingly vital as these quantities increase. This is especially true for network data, where, in its simplest form, the size of a data set increases quadratically with the sample size. Alternatively, the approach may also prove useful for more sophisticated models with additional parameters to be estimated. For example, in the case of continuous data, a mixture of skewed, heavy-tailed distributions such as the skew-t \citep{McLachlan1998,Andrews2011,Lee2012} may be preferred to a mixture of Gaussians.

The BIA method has been motivated by the success of Bayesian model averaging in many applications~\citep{Hoeting99}. It might be argued that the use of a BIC-like penalty places too much weight on small models. While we suggest that the use of this penalty term is attractive based on theoretical grounds, it would be interesting to see whether the use of different penalty terms, such as the milder AIC term, yielded superior results in some settings.

An increased number of EM runs and starting positions will generally boost performance of the BIA approach, but as in the case of tempering \citep{zhou2010} there is no single systematic rule for selecting an optimum combination, and some trial and error is required. However, extensive experimentation on the motivating data sets has shown that the results presented are robust for wide ranges of number of EM runs and number of starting positions. Intuitively it seems reasonable to expect that, if only one of these two inputs is to be elevated, one should favour many starting positions with a relatively small number of runs for each, since the EM algorithm usually increases likelihood most substantially in the early iterations. For the data sets analyzed, this intuition holds for the \emph{Karate}, \emph{AIS} and \emph{Carcinoma} data, but not the \emph{Alzheimer's} data.

In the examples shown, different starting values were generated using standard approaches. This was done to ensure simple and fair settings were used to compare BIA with competing initialization methods such as random starts, hierarchical clustering, and deterministic annealing. However, this does not have to be the case. For example, in the case of continuous data, clustering solutions obtained from the different covariance structures available in \textbf{mclust} could be used. An advantage of the BIA approach is that rather than discarding competing heuristic approaches, they may be assimilated in a principled, statistically robust manner that efficiently harnesses the available computational resources.

%

%

\begin{acknowledgements}
The authors would like to acknowledge the contribution of Dr Jason Wyse to this paper, who provided many helpful insights as well as C++ code for the label-switching methodology employed.
\end{acknowledgements}

\bibliographystyle{spbasic}      
\bibliography{CSbibOHaganandWhite}   

%
%

\end{document}